\documentclass[iop]{emulateapj}
\usepackage{bm}
\usepackage{threeparttable}
\usepackage{morefloats}
\usepackage{CJK}
\def\msun{M_\odot}
\def\mbh{M_{\rm{BH}}}
\def\Medd{\dot{M}_{\rm{Edd}}}
\def\Ledd{L_{\rm{Edd}}}

\newcommand\bF{{\mbox{\boldmath $F$}}}

\def\<{\,\langle\langle}
\def\>{\,\rangle\rangle}

\begin{document}
\begin{CJK*}{UTF8}{gbsn}

\shortauthors{Y.-F. Jiang et al.}
\author{Yan-Fei Jiang(姜燕飞)\altaffilmark{1}, Omer Blaes\altaffilmark{2}, James M. Stone\altaffilmark{3} \& Shane W. Davis\altaffilmark{4}}
%\author{Yan-Fei Jiang\altaffilmark{1,2}\footnote{Einstein Fellow}, James M. Stone\altaffilmark{1} \& Shane W. Davis\altaffilmark{3}}
\affil{$^1$kavli institute for theoretical physics, Kohn Hall, University of California, Santa Barbara 93106, USA} 
\affil{$^2$Department of Physics, University of California, Santa Barbara, CA 93106, USA}
\affil{$^3$Department of Astrophysical Sciences, Princeton
University, Princeton, NJ 08544, USA} 
\affil{$^4$Department of Astronomy, University of Virginia, P.O. Box 400325, Charlottesville, VA 22904-4325, USA}

\title{Global Radiation Magneto-hydrodynamic Simulations of sub-Eddington Accretion Disks around Supermassive black Holes}

\begin{abstract}
We use global three dimensional radiation magneto-hydrodynamic simulations to study the properties of inner regions of accretion disks around a $5\times 10^8\mbh$ black hole with mass accretion rates reaching $7\%$ and $20\%$ of the Eddington value. 
%$0.7\Ledd/c^2$ and $2\Ledd/c^2$. 
This region of the disk is supported by magnetic pressure with surface density significantly smaller than the values predicted by the standard thin disk model but with a much larger disk scale height. The disks do not show any sign of thermal instability over many thermal time scales. More than half of the accretion is driven by radiation viscosity in the optically thin corona region for the lower accretion rate case, while accretion in the optically thick part of the disk is driven by the Maxwell and Reynolds stresses from MRI turbulence. Coronae with gas temperatures $\gtrsim 10^8{\rm K}$ are generated only in the inner $\approx 10$ gravitational radii in both simulations, being more compact in the higher accretion rate case. In contrast to the thin disk model, surface density increases with increasing mass accretion rate, which causes less dissipation in the optically thin region and a relatively weaker corona. 
The simulation results may explain the formation of X-ray coronae in Active Galactic Nuclei (AGNs), the compact size of such coronae, and the observed
trend of optical to X-ray luminosity with Eddington ratio for many AGNs.

\end{abstract}


\maketitle

\section{Introduction}
Active Galactic Nuclei (AGNs) are believed to be powered by accretion onto the central black holes. Luminosity of most AGNs is observed to be smaller than the Eddington limit as defined by the electron scattering opacity $\Ledd=1.5\times 10^{46}\mbh/\left(10^8\msun\right)$ erg s$^{-1}$. The standard thin disk model (\citealt{ShakuraSunyaev1973}) is usually adopted to describe the accretion disks in this regime when the AGN luminosity is larger than $\approx 0.01\Ledd$. The key assumptions in this model are that the disk is optically thick and supported by thermal pressure (gas or radiation), and all the dissipation is radiated away locally. The detailed structures of the disk are then determined by the famous $\alpha$ assumption, which assumes that the stress responsible for the angular momentum transport in the disk is proportional to the total pressure with a constant value. 

Comparing the predictions from the standard thin model with the observed properties of AGNs has raised many questions on the assumptions in this model (\citealt{KoratkarBlaes1999}). The standard thin disk model subjects to the well known thermal and inflow instabilities (\citealt{ShakuraSunyaev1976,lightmanEardley1974,Piran1978}), which should cause the disk to evolve away from the equilibrium state in a few thermal time scales (\citealt{Jiangetal2013c,Fragileetal2018}) and probably show some kind of limit-cycle behavior (\citealt{Honmaetal1991,Janiuketal2002}). However, this is not observed for most AGNs. The predicted spectrum from this model is also not consistent with the observed energy distribution (\citealt{Zhengetal1997,Davisetal2007,LaorDavis2014}). Particularly, the predicted edge feature in the spectrum is also not observed (\citealt{SincellKrolik1997,Shulletal2012,Tiltonetal2016}). 

The standard thin disk model assumes that thermal pressure (gas or radiation) is dominant in the disk. Alternative possibility that the vertical component of gravity in the accretion disk may be primarily balanced by magnetic pressure has been proposed (\citealt{Shibataetal1990,Parievetal2003}). \cite{BegelmanPringle2007} studied the structures and stability of magnetic pressure dominated disks by assuming the toroidal magnetic fields will saturate to a level so that the associated Alfv\'en velocity reaches $\sqrt{c_sV_k}$ (\citealt{PessahPsaltis2005,Dasetal2018}), where $c_s$ is the gas sound speed and $V_k$ is the disk rotation speed. Amplification of magnetic fields near the disk midplane is thought to be balanced by the escape of magnetic fields away from the midplane due to buoyancy. This magnetic elevated model is found to have a larger pressure scale height and does not subject to the thermal and viscous instabilities (\citealt{Sadowski2016}), which have interesting implications for both X-ray binaries and AGNs (\citealt{Begelmanetal2015,BegelmanSilk2017,DexterBegelman2019}). However, assumptions in the magnetic elevated disk model have not been checked numerically. It is also unclear how the strong radiation pressure in AGN accretion disks will modify the structures, which is typically neglected in these models. It is known that in order to reach the magnetic pressure dominated state, large scale poloidal or radial magnetic fields are required (\citealt{Salvesenetal2016,Salvesenetal2016b,FragileSadowski2017}). The advantage of our simulations is that gas, radiation and  magnetic pressure are all included self-consistently.  Our goal is to study the structures of the disk that will be formed with these magnetic field configurations for realistic parameters of AGNs. Our simulations form the a disk in a region that is initially vacuum by accreting gas from larger radius. Although the resulting accretion disk depends on our initial condition for the magnetic fields, the disk forms self-consistently. This distinguishes it from previous work that initialized the simulation with a standard thin disk model and study its long term evolution (\citealt{Fragileetal2018}).

Formation of coronae in AGNs, as well as the dependence of corona properties on the accretion flow, remain a puzzle. The effective temperature of the accretion disks in AGNs (typically $\sim 10^4-10^5$ K) is too low to produce the widely observed X-rays in AGNs via thermal emission. 
It is proposed that high temperature corona, as inspired by the solar corona, is formed on top of the accretion disk (\citealt{BisnovatyiBlinnikov1976,HaardtMaraschi1991,HaardtMaraschi1993,SvenssonZdziarski1994,Zdziarskietal1999}). X-rays are produced via Compton scattering of the seed photons emitted by the disk with the hot electrons in the corona, which is infereed to exist in a compact region near the black hole ($\sim 10$ gravitational radii, \citealt{ReisMiller2013,Uttleyetal2014}). The amount of X-rays that can be produced via this mechanism compared with the thermal emission from the disk is determined by the fraction of accretion power that is dissipated in the corona region. This is basically a free parameter in these models since the mechanism to produce the corona in AGNs is unknown. Since the early isothermal simulations of vertically stratified accretion disk with turbulence generated by the Magneto-rotational instability (MRI) (\citealt{MillerStone2000}), it is commonly observed that magnetic fields amplified by MRI near the midplane of the disk can buoyantly rise to the surface and create a magnetic pressure dominated low density region, which is suspected to be the corona. However, most local simulations that determine the thermal properties of the disks based on radiative cooling (using both the flux-limited diffusion and VET method) typically find that the gas in the magnetic pressure dominated region is not heated to a very high temperature due to insignificant dissipation (\citealt{Kroliketal2007,Blaesetal2007,Blaesetal2011,Hiroseetal2009,Jiangetal2013c,Jiangetal2016a}). An exception is \cite{Jiangetal2014}, which shows that the amount of dissipation in the magnetic pressure dominated region can be increased by reducing the surface density of the disk, which can increase the gas temperature in this region. Surface density is a free parameter in these local shearing box simulations and the value required to produce the high temperature corona is smaller than what the thin disk model will predict for the same accretion rate. It is therefore necessary to use global simulations to determine the surface density of the disk as well as the properties of coronae self-consistently for a given mass accretion rate, which is one goal of the paper.

\cite{Jiangetal2016a} shows that for typical density and temperature expected for AGN accretion disks based on the standard thin disk model, the Rosseland mean opacity should be larger than the electron scattering value due to irons. The density and temperature dependences of the iron opacity peak can modify the thermal stability and structures of AGN accretion disks significantly. However, the local shearing box simulations done by \cite{Jiangetal2016a} adopt the surface density as given by the standard thin disk model. 
Whether this opacity peak will show up or not depends on the actual disk structure we will get. We will include the full opacity table in the simulations to capture its potential importance. 

The remainder of this paper is organized as follows. In
Section \ref{sec:setup}, we describe the simulation setup. 
Detailed structures of the disk are described in Section \ref{sec:result}.
We discuss the implications of our simulations in Section \ref{sec:discussion}.

\section{Simulation Setup}
\label{sec:setup}
We solve the same set of ideal  MHD equations coupled with the time dependent radiative transfer equation for specific intensities as in \cite{Jiangetal2018} using the code {\sf Athena++} (Stone et al, in preparation). We carry out two simulations, {\sf AGN0.2} and {\sf AGN0.07}, for a $\mbh=5\times 10^8\msun$ black hole with accretion rates smaller than the Eddington value $\Medd\equiv 10\Ledd/c^2=8.22\times 10^{26}\ {\rm g/s}$. The simulations are performed with the pseudo-Newtonian 
potential \citep[][]{PaczynskiWiita1980} $\phi=-G\mbh/(r-2r_g)$ to mimic the general relativity effects around a Schwarzschild black hole, where $G$ is the gravitational constant while gravitational radius $r_g\equiv G\mbh/c^2=7.42\times 10^{13}\ {\rm cm}$. We initialize a torus  centered at $80r_g$ with the maximum density $\rho_0=10^{-8}\ {\rm g}\ {\rm cm}^{-3}$ and temperature $3.18T_0$, where $T_0=2\times 10^5 {\rm K}$. The shape of the torus is the same as the ones used in \cite{Jiangetal2018}. The inner edge of the torus is at $40r_g$ and the region inside that radius is filled with density floor $10^{-9}\rho_0$ initially. The initial properties of the torus, including the ratios between the averaged radiation pressure $P_r$, gas pressure $P_g$ and magnetic pressure $P_B$, are summarized in Table \ref{Table:parameters}. The main difference between the two simulations is the initial magnetic field in the torus. The run {\sf AGN0.2} uses a single loop of poloidal magnetic field while the run {\sf AGN0.07} adopts multiple magnetic field loops as described in \cite{Jiangetal2018}. The different setups lead to different mass accretion rates in the disks that are formed near the black hole. 
    
We use four levels of static mesh refinement to cover the whole simulation domain $\left(r,\theta,\phi\right)\in\left(4r_g, 1600r_g\right)\times \left(0,\pi\right)\times \left(0,2\pi\right)$. The level with the highest resolution reaches $\Delta r/r=\Delta \theta=\Delta\phi=6.1\times 10^{-3}$ for the region $\left(6r_g,200r_g\right)\times \left(1.48,1.66\right)\times\left(0,2\pi\right)$, which covers most of the mass near the disk midplane. The equivalent  resolution is $1024\times 512\times 1024$, which is necessary to resolve these sub-Eddington accretion disks. 

We use $80$ discrete angles in each cell to resolve the angular distribution of the radiation field, which is in thermal equilibrium initially in the torus.
% To capture the potential importance of the iron opacity bump in AGN accretion disks %(\citealt{Jiangetal2016a}), 
We calculate the Rosseland mean opacity in each cell by using local density and temperature based on the OPAL opacity table with solar metallicity (\citealt{Paxtonetal2013,Jiangetal2015}). Planck mean free-free absorption opacity  is also included as in \cite{Jiangetal2018}.  Each simulation takes $\approx 20-30$ millions CPU time in the ALCF machine {\sf Mira}.

\begin{table*}[h]
	\caption{Simulation Parameters}
	\begin{center}
		\begin{tabular}{ccccc}
			\hline
			Variables/Units & {\sf AGN0.2} & {\sf AGN0.07} \\
			% This corresponds to run AGNGlobal3 and AGN3B
			\hline
			$r_i/r_g$              		      		&	  80	    & 	 80		\\	
			$\rho_i/\rho_0$			      		&	  1    &       1 	\\
			$T_i/T_0$                               &   3.18       &    3.18   \\
			$ \langle P_{r}/ P_{g}\rangle$		& $4.47\times 10^5$	    &  	  $4.61\times 10^5$ 	\\
			$ \langle P_{r}/P_{g}\rangle_{\rho}$   &  $4.40\times 10^2$   &   $4.39\times 10^2$      \\
			$ \langle P_{B}/P_{g}\rangle$		 &	$1.27\times 10^{-2}$    &  	 $2.32\times 10^{-4}$	\\
			$ \langle P_{B}/P_{g}\rangle_{\rho}$   &   $7.70\times 10^{-3}$  &   $7.32\times 10^{-5}$        \\
			$\Delta r/r$                               		 & $6.1\times 10^{-3}$  &	 $6.1\times 10^{-3}$		\\
			$\Delta \theta$		               		&   $6.1\times 10^{-3}$ &   $6.1\times 10^{-3}$	 \\
			$\Delta \phi$	                       		&  $6.1\times 10^{-3}$  &  $6.1\times 10^{-3}$		\\
			$N_n$                                     		&  80    &  80  \\
			\hline
		\end{tabular}
	\end{center}
	\label{Table:parameters}
	\begin{tablenotes}
		\item Note: The center of the initial torus is located at $r_i$ with density and temperature to be $\rho_i$ and $T_i$. 
		The fiducial density and temperature are $\rho_0=10^{-8}\ {\rm g}\ {\rm cm}^{-3}$ and $T_0=2\times 10^5 {\rm K}$.
		For any quantity $a$, $\langle a \rangle$ is the volume averaged value over the torus while $\langle a \rangle_{\rho}$ 
		is the averaged value weighted by the mass in each cell. The grid sizes   
		$\Delta r, \Delta \theta, \Delta\phi$ are for the finest level at the center of the torus. The number 
		of angles for the radiation grid is $N_n$ in each cell. 
	\end{tablenotes}
\end{table*}

\section{Results}
\label{sec:result}
\subsection{Resolution for MRI}
To quantify how well the MRI is resolved in our simulations, we calculate the quality factors $Q_{\theta}$ and $Q_{\phi}$, which are the ratios between the wavelength of the fastest growing MRI mode $\lambda=2\pi\sqrt{16/15}|v_{A}|/\Omega$ and cell sizes $r\Delta \theta, r\sin\theta\Delta \phi$ along $\theta$ and $\phi$ directions, where the Alfv\'en velocity $v_{A}$ is calculated for $B_{\theta}$ and $B_{\phi}$ respectively. Resolution studies for non-radiative ideal MHD simulations \citep{Hawleyetal2011,Sorathiaetal2012} find that when $Q_{\phi}  \gtrapprox 25, Q_{\theta}\gtrapprox 6$ or both $Q_{\phi}$ and $Q_{\theta}$ are larger than $10$, properties of MRI turbulence is converged with respect to resolution. Although there is no general criterion on the convergence of radiation MHD simulations of accretion disks, we use it as a way to compare our resolutions with these calculations.  We calculate the azimuthally averaged quality factors during the final turbulent states for both of the calculations. 
For the simulation {\sf AGN0.07} at the disk midplane, $Q_{\phi}=100$ at $r=5r_g$ and decreases to 80 at $r=30r_g$, while $Q_{\theta}$ varies from 10 to 8 in the same radial range. For a fixed radius, the combined effects of increasing Alfv\'en velocity and reduced resolution with height from the disk midplane cause the quality factors to decrease by a factor of $\approx 2$ at the photosphere. For the simulation {\sf AGN0.2}, the quality factors are larger as the disk is thicker. The midplane $Q_{\phi}$ changes from 200 at $5r_g$ to 40 at $40r_g$ while $Q_{\theta}$ changes from $\approx 20$ to $\approx 10$ between $5$ and $40r_g$ at the disk midplane. For this run, the quality factors increase by a factor of $\approx 2$ from the midplane to the photosphere. Therefore, the resolutions are sufficient to resolve the MRI turbulence according to the criterion found by non-radiative MHD simulations. This is possible for these sub-Eddington accretion disks because magnetic pressure is the dominant pressure, which makes the wavelength of the fastest growing MRI mode be comparable to the disk scale height and easier to resolve.

%The general concern for simulating accretion disks in the sub-Eddington %regime is that wavelength of the fastest growing MRI mode is too small %compared with the radius to be resolved if the disk is dominated by the %thermal pressure as in the standard thin disk model. 
%However, this is not the case 

\begin{figure*}[htp]
	\centering
	\includegraphics[width=0.49\hsize]{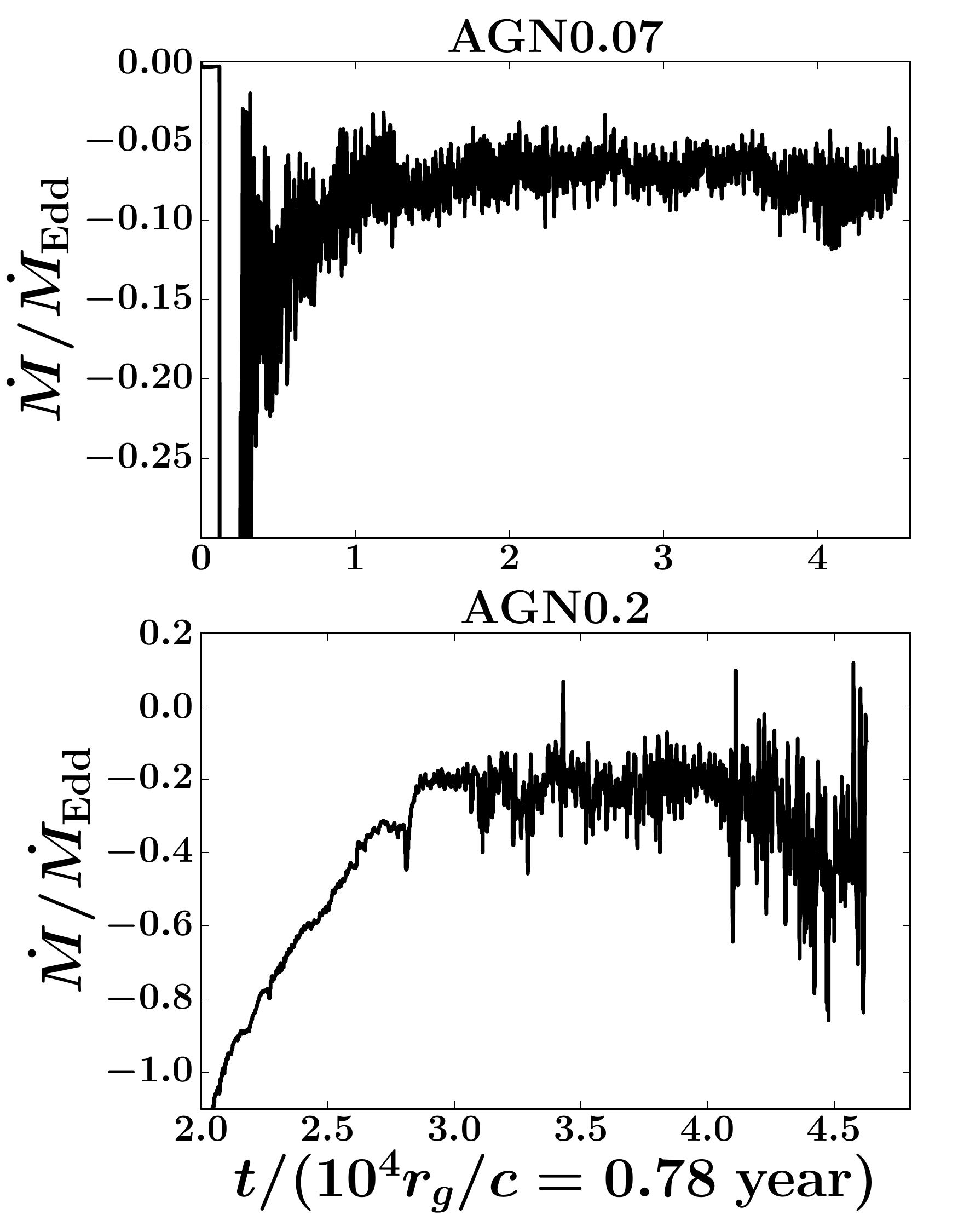}
	\includegraphics[width=0.49\hsize]{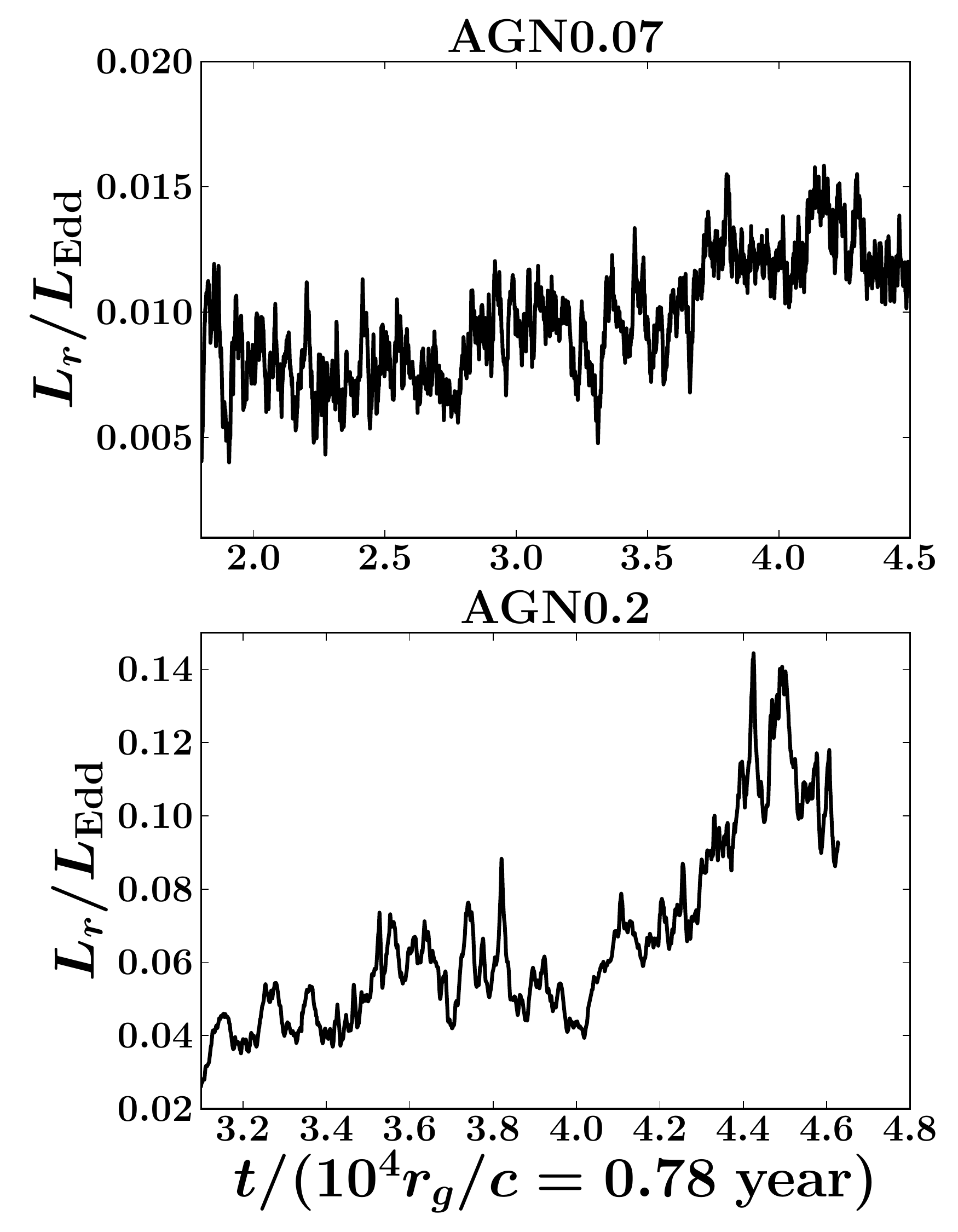}
	\caption{\emph{Left:} histories of the net mass accretion rate (negative values for inflow) at $10r_g$ for the two simulations 
		{\sf AGN0.07} (top panel) and {\sf AGN0.2} (bottom panel). The total durations of the simulations correspond to more than 10 thermal time scales at $10r_g$. \emph{Right:} histories of total radiative luminosity emitted within the cylindrical radius $10r_g$ of the disk for simulations {\sf AGN0.07} (top panel) and {\sf AGN0.2} (bottom panel). Note that the horizontal axes have different offsets in the panels.}
	%		 The time 
	%		unit $t_0\equiv r_g/c$ corresponds to $5\times10^{-3}$ Keplerian orbital period at $10r_g$.}
	\label{mdot_hist}
\end{figure*}

\begin{figure*}[htp]
	\centering
	\includegraphics[width=0.49\hsize]{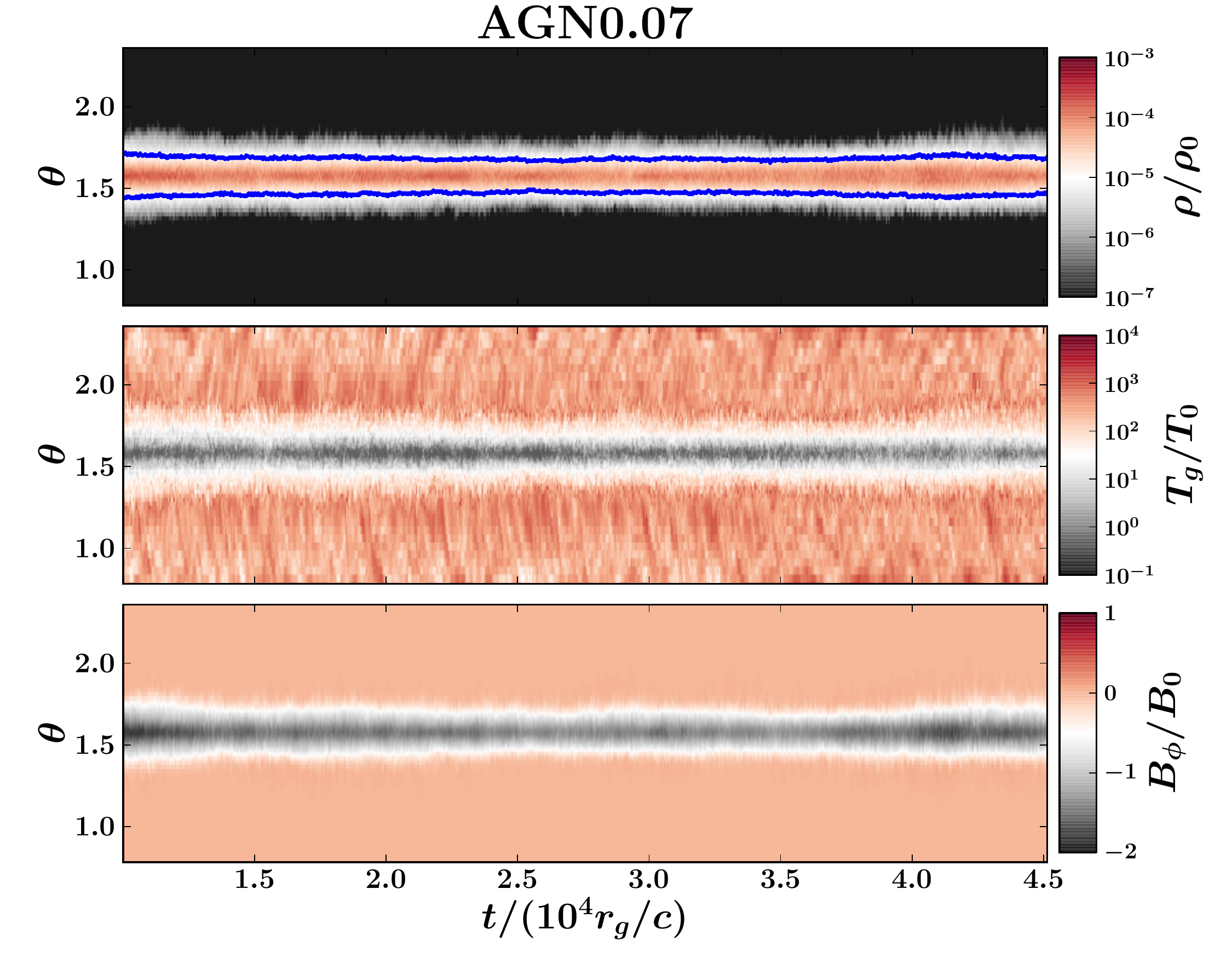}
	\includegraphics[width=0.49\hsize]{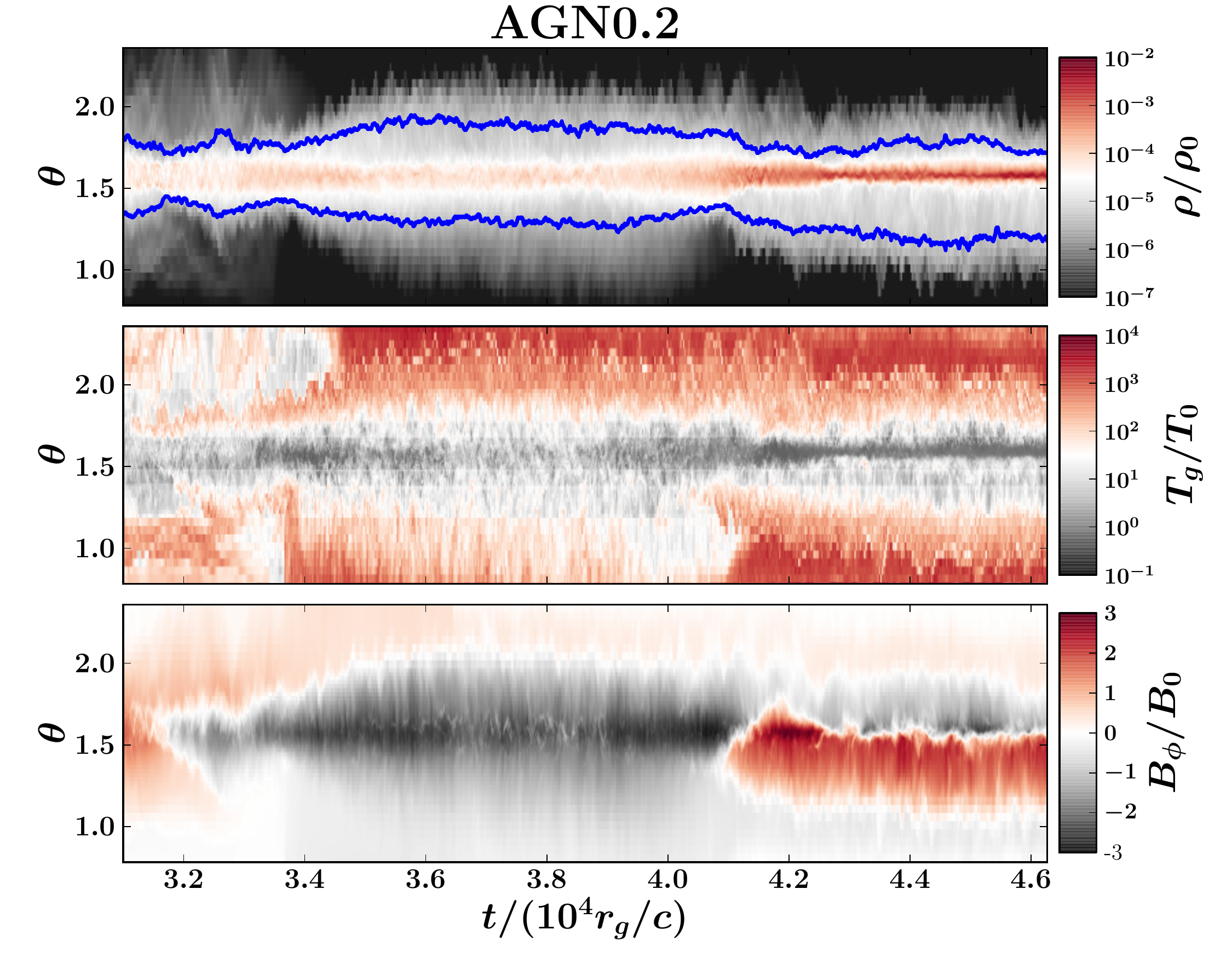}
	\caption{
		Space-time diagrams for the histories of azimuthally averaged density ($\rho$ in unit of $\rho_0=10^{-8}$ g/cm$^3$, top panels), gas temperature ($T_g$ in unit of $T_0=2\times 10^5$ K, middle panels) and toroidal magnetic field ($B_{\phi}$ in unit of $B_0=1.87\times 10^3$ G, bottom panels) at $10r_g$. The two columns are for the run {\sf AGN0.07} (left) and {\sf AGN0.2} (right) respectively. The blue lines in the top panel indicate the location where optical depth for Rosseland mean opacity to the rotation axis is one. }
	\label{STplot}
\end{figure*}

\begin{figure}[htp]
	\centering
	\includegraphics[width=1.0\hsize]{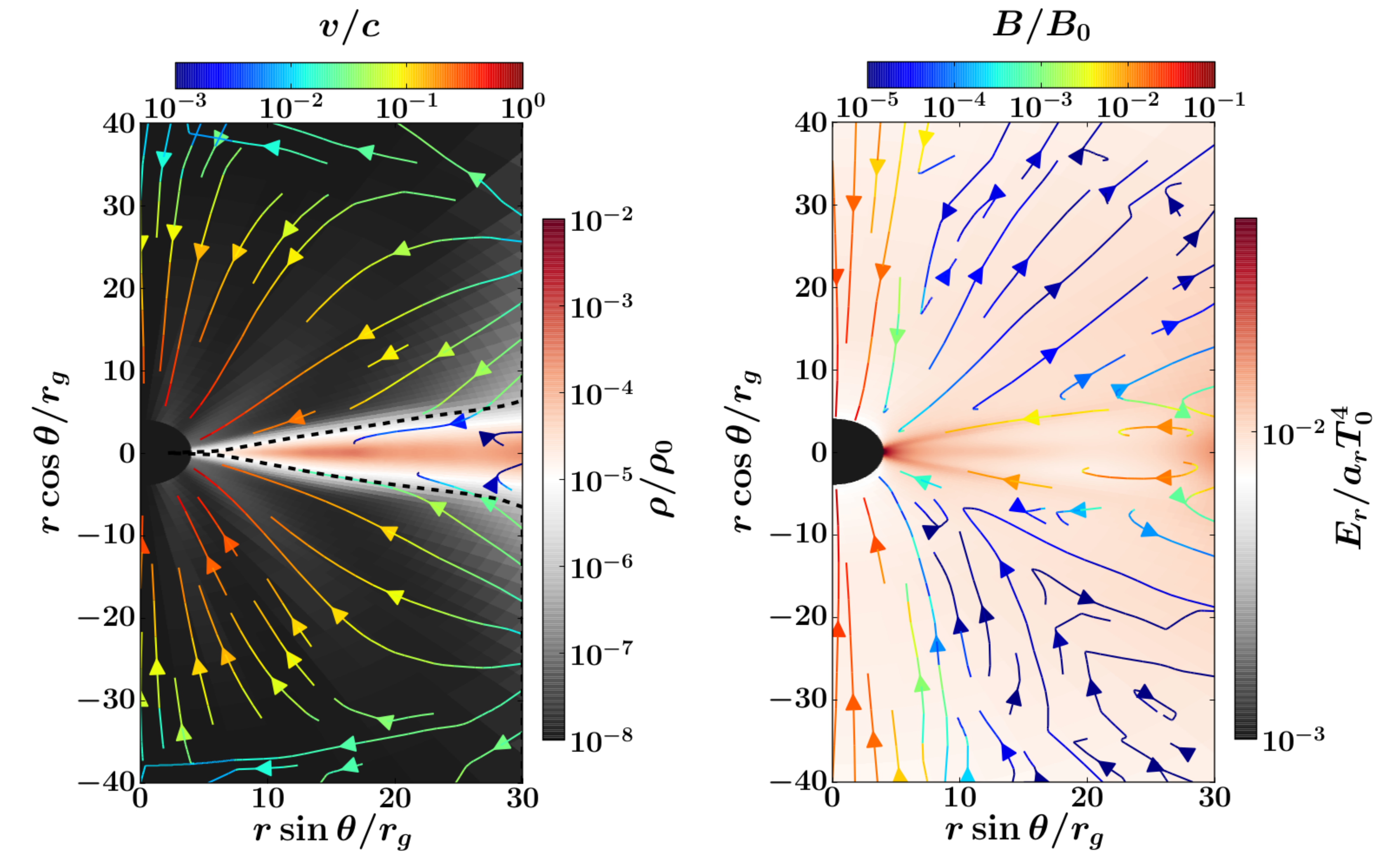}
	\includegraphics[width=1.0\hsize]{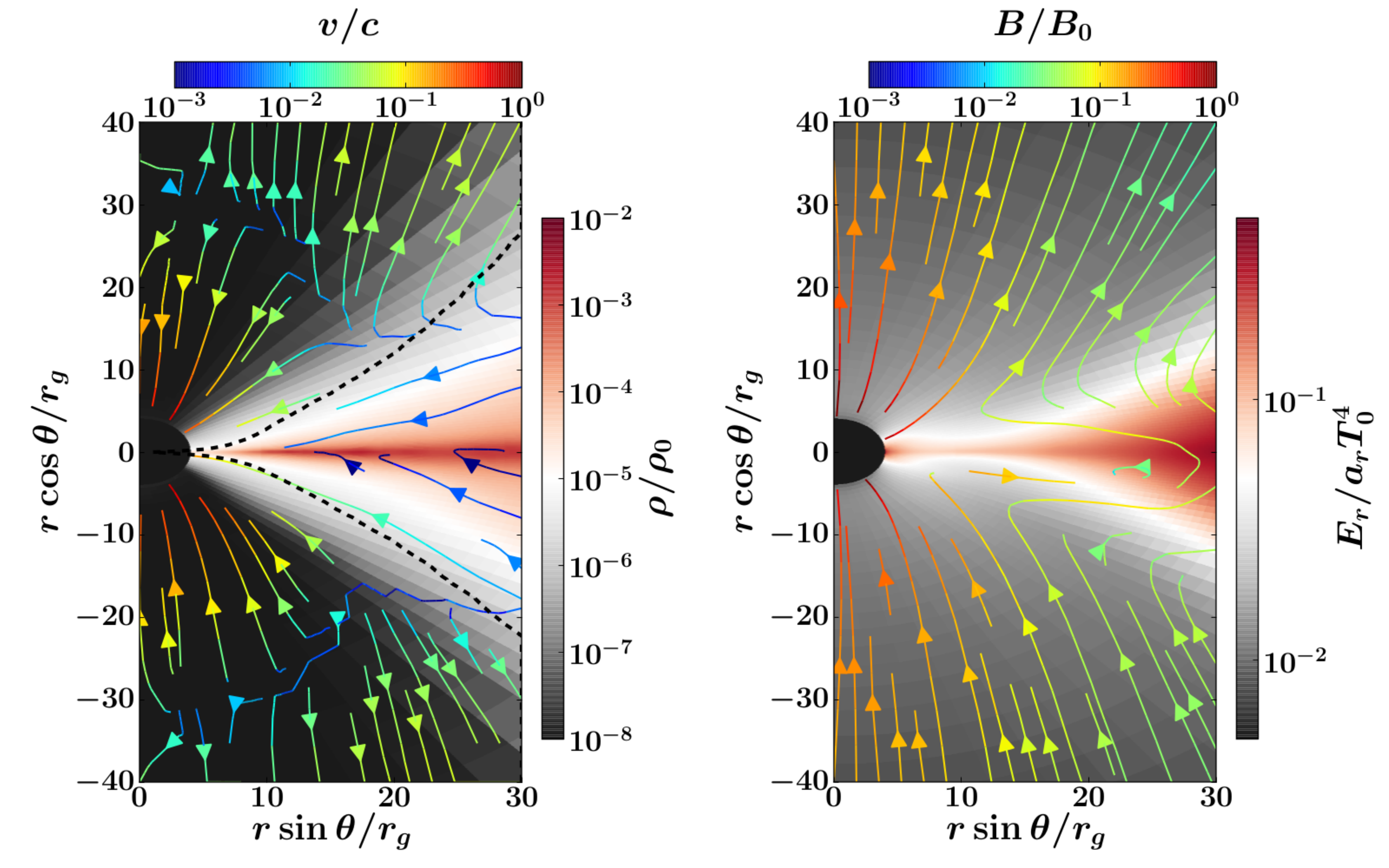}
	\caption{Time and azimuthally averaged spatial structures for the inner $40r_g$ of the disks from the two simulations {\sf AGN0.07} (top panels) and {\sf AGN0.2} (bottom panels). The left column is for density (the color) and flow velocity (the streamlines) while the right column is for radiation energy density (the color) and magnetic field lines (the streamlines). The dashed black line in the left column indicates the location where optical depth to the rotation axis for Rosseland mean opacity is $1$.}
	%		 The time 
	%		unit $t_0\equiv r_g/c$ corresponds to $5\times10^{-3}$ Keplerian orbital period at $10r_g$.}
	\label{ave_disk_structure}
\end{figure}

\subsection{Simulation Histories}
Gas in the initial torus flows towards the black hole and the accretion disk is slowly built up. Histories of the mass accretion rate $\dot{M}\equiv \int_0^{2\pi}\int_0^{\pi}\rho v_r r^2\sin\theta d\theta d\phi$ at $10r_g$ for the two runs are shown in the left panel of Figure \ref{mdot_hist}. After a period of initial transient, the simulation {\sf AGN0.07} reaches an accretion rate $7\%\Medd$ when averaged after time $t=2.4\times 10^4r_g/c$, which roughly corresponds to $13$ thermal time scales at $10r_g$. 
For the run {\sf AGN0.2}, the averaged accretion rate is $22\%\Medd$ after $t=3\times 10^4r_g/c$, which also corresponds to more than $\approx 6$ thermal time scales at $10r_g$. All the analysis for the time averaged properties of the two simulations are also performed for the same time intervals. The mass accretion rates show significant fluctuations due to turbulence.  The standard deviation of $\dot{M}$ during the steady state for the run {\sf AGN0.07} is only $1.2\%$ of the averaged value while the corresponding value is $12.3\%$ for the run {\sf AGN0.2}. The main difference of the two simulations is the magnetic field topology as described in Section \ref{sec:setup}. The run {\sf AGN0.2} adopts a single loop of magnetic field in the initial torus and there are net poloidal magnetic fields $\overline{B}_{\theta}$ through the inner region of the disk. 
The ratio between radiation pressure and $\overline{B}_{\theta}^2/2$ near the disk midplane is $\approx 10^3$. For the run {\sf AGN0.07}, quadrupole magnetic fields are used in the initial condition, which result in net radial magnetic fields $\overline{B}_r$ near the disk midplane, although the shell averaged $B_r$ across the whole disk is still zero. The net $\overline{B}_r$ near the disk midplane is sheared into toroidal magnetic field within an orbital time scale and quickly builds up strong magnetic pressure, which then escapes from the midplane due to buoyancy. MRI turbulence is still developed in this case (\citealt{PessahPsaltis2005,Dasetal2018}), but it shows less variability compared with the other run. 
Detailed investigations of magnetic pressure dominated disks resulting from this magnetic field configuration are described in Section \ref{sec:ave_structure}.
%The Maxwell stress from the turbulent component is still order of magnitude larger than the stress %due to the mean magnetic field $-\overline{B}_r\overline{B}_{\phi}$ (see Section %\ref{sec:ave_structure}), where $\overline{B}_{\phi}$ is the azimuthally averaged toroidal magnetic %field. The ratio between radiation pressure and $\overline{B}_r^2/2$ near the disk midplane during %the steady state varies from $100$ at $8r_g$ to $10^4$ at $15r_g$. 

In most MHD simulations of accretion disks where radiative transfer is not calculated self-consistently, density is usually used as the proxy to understand the observed properties of these systems. Since our simulations calculate the photons emitted by the disk directly, we can calculate the frequency-integrated lightcurves for the time scale that our simulations can cover. In order to avoid contamination from photons generated by the torus at large radii where the disk has not reached steady state, we convert the radiation flux ${\bF_r}$ to the radial $F_R$ and vertical components $F_z$ in cylindrical coordinates and then integrate the total radiative luminosity leaving the cylindrical surface at $R=10r_g$ and height $z=100r_g$, which is well beyond the photosphere. The resulting lightcurves are shown in the right column of Figure \ref{mdot_hist}, which have much weaker short time scale variabilities compared with what the variability in $\dot{M}$ might suggest. This is likely due to the scattering of photons through the optically thick disks. 
In particular, the power spectrum of the lightcurve from the lower accretion rate run ({\sf AGN0.07}) has more power at high frequencies compared with the lightcurve power spectrum from the run with higher accretion rate ({\sf AGN0.2}). This is because the midplane optical depth increases with increasing mass accretion rate (see Figure~\ref{surface_density}).
% Due to the limited radial range for which we can achieve steady state in these simulations, %these lightcurves can only be compared with the X-ray emission from AGNs. (MORE DISCUSSIONS %here.)
The changes of variability amplitude with luminosity are also different in the two lightcurves, revealing different dynamo actions in the disk, which will be discussed in Section \ref{sec:discussion}.

%They are consistent with the general observed trend that variability of AGN is larger with lower %mass accretion rate. 

Histories of poloidal profiles of azimuthally averaged density $\rho$, gas temperature $T_g$ and toroidal magnetic field $B_{\phi}$ at $10r_g$ for the two runs are shown in Figure \ref{STplot}. The disk photosphere only extends to $\approx 6^{\circ}$ away from the disk midplane for the run {\sf AGN0.07} while in the simulation {\sf AGN0.2}, the disk photosphere covers more than $14^{\circ}$. Once above the photosphere as indicated by the blue lines in the top panels, gas temperature increases rapidly to $\approx 10^9$ K, which is the corona region. More detailed properties of the corona are discussed in Section \ref{sec:corona}. In the optically thick part of the disk, gas and radiation are in thermal equilibrium with a temperature around $10^5$ K. 
The grid scale variation of gas temperature is unrealistic because gas pressure is smaller than $0.1\%$ of radiation and magnetic pressure and it suffers from numerical noise. The toroidal magnetic field switches signs near the disk midplane after $\approx 10^4r_g/c$ for the run {\sf AGN0.2}, which is the well known butterfly diagram caused by the MRI dynamo \citep{Stoneetal1996,MillerStone2000,Davisetal2010,ONeilletal2011,Simonetal2012,Jiangetal2013c,Jiangetal2014c}. This does not happen for the run {\sf AGN0.07}, although MRI turbulence has also developed there. This is because shearing of the net radial magnetic field near the disk midplane for the run {\sf AGN0.07} always forms $B_{\phi}$ with the same sign in addition to the $B_{\phi}$ generated by MRI (see Section \ref{sec:vertical}).

\begin{figure}[htp]
	\centering
	\includegraphics[width=1.0\hsize]{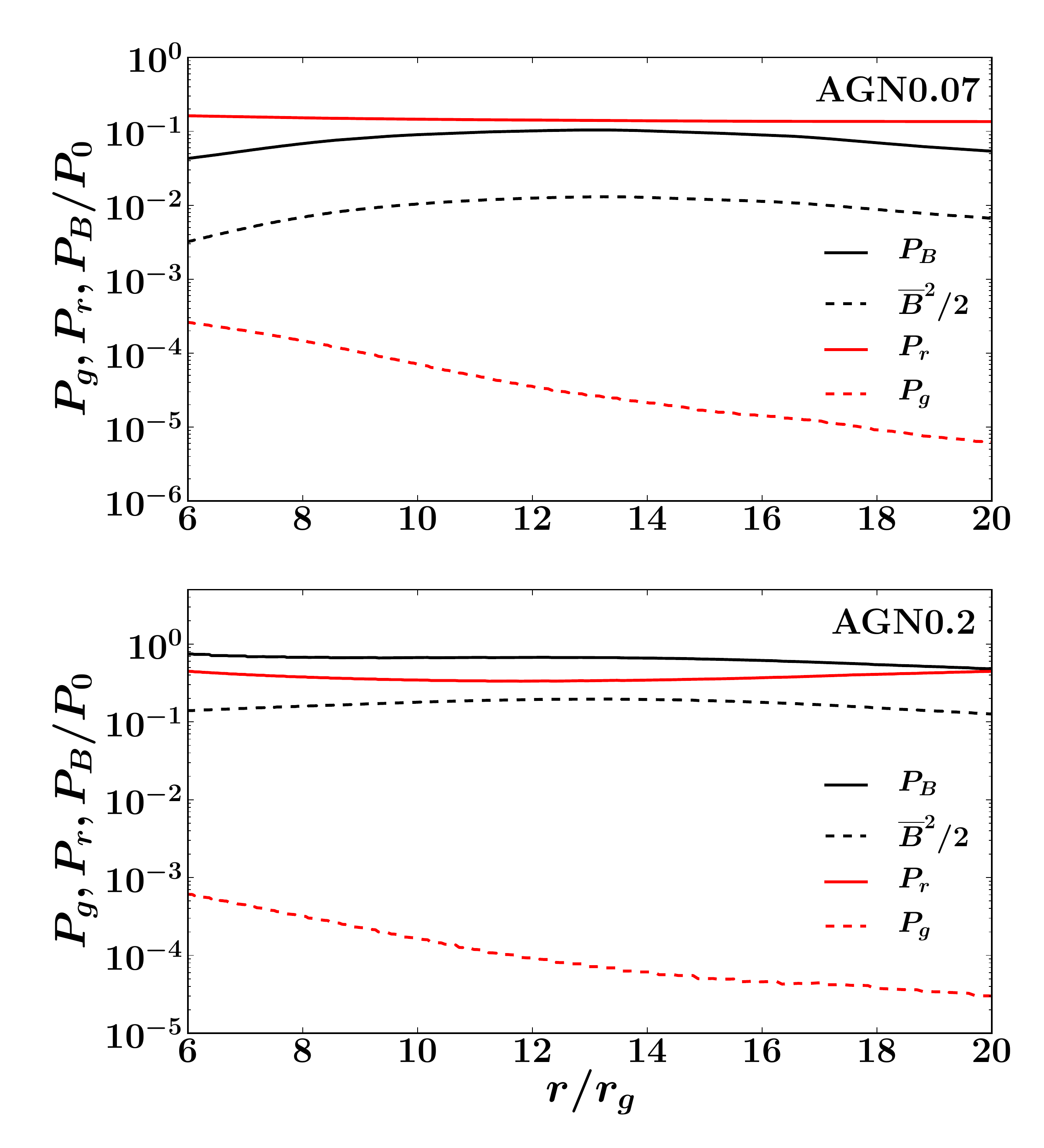}
	\caption{Radial profiles of the time and shell averaged gas pressure ($P_g$, dashed red lines), radiation pressure ($P_r$, solid red lines), total magnetic pressure ($P_B$, solid black lines) and magnetic pressure due to the azimuthally averaged magnetic field ($\overline{B}^2/2$, dashed black lines) in the two simulations {\sf AGN0.07} (top panel) and {\sf AGN0.2} (bottom panel). }
	\label{pressure_profile}
\end{figure}

\begin{figure}[htp]
	\centering
	\includegraphics[width=1.0\hsize]{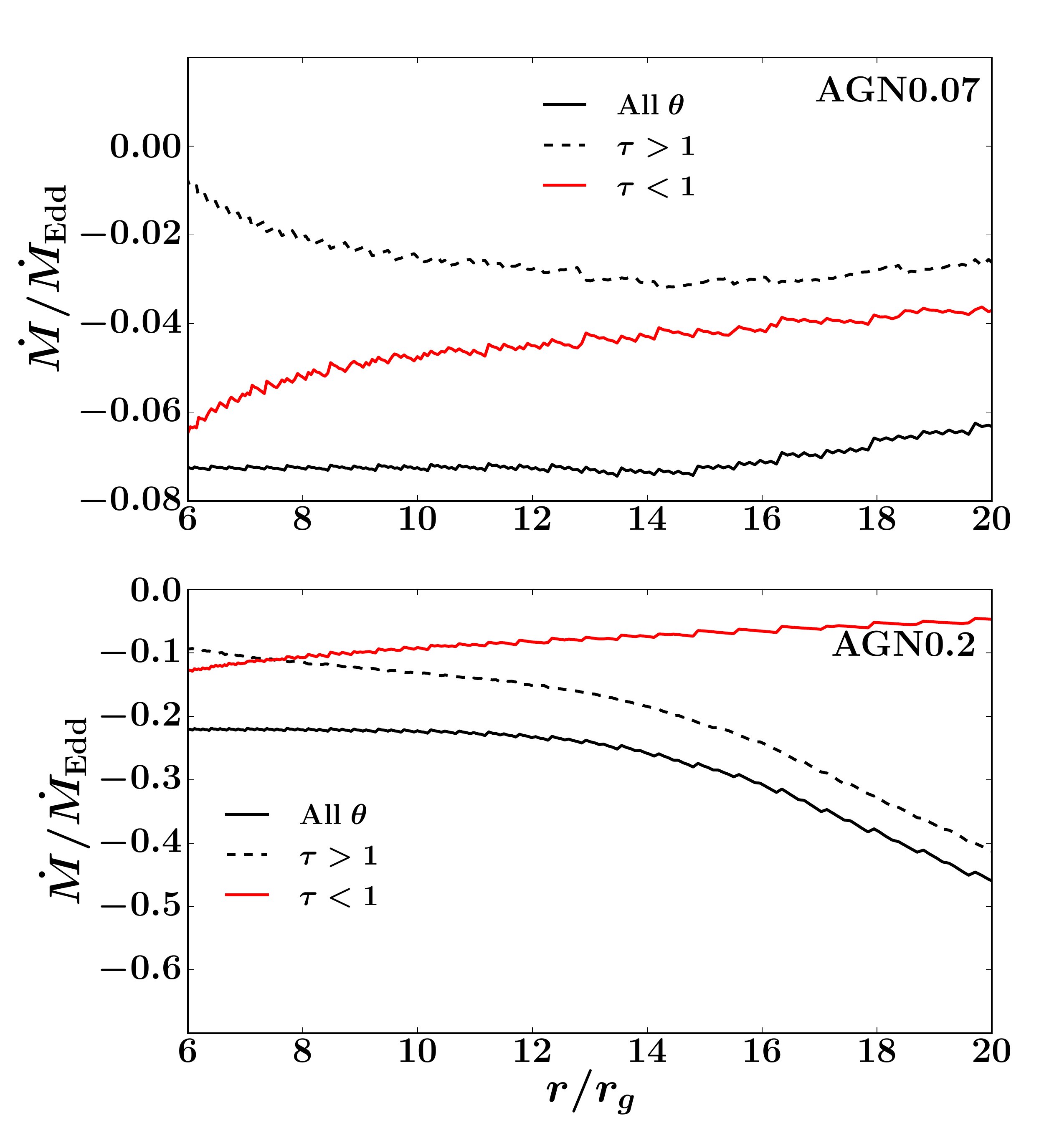}
	\caption{Radial profiles of the time averaged net mass accretion rate $\dot{M}$ for the two runs {\sf AGN0.07} (top panel) and {\sf AGN0.2} (bottom panel). The solid black lines are $\dot{M}$ integrated over all polar angles, while the solid red and dotted black lines are $\dot{M}$ in the optically thin and thick regions respectively.}
	\label{Mdot_profile}
\end{figure}

\begin{figure}[htp]
	\centering
	\includegraphics[width=1.0\hsize]{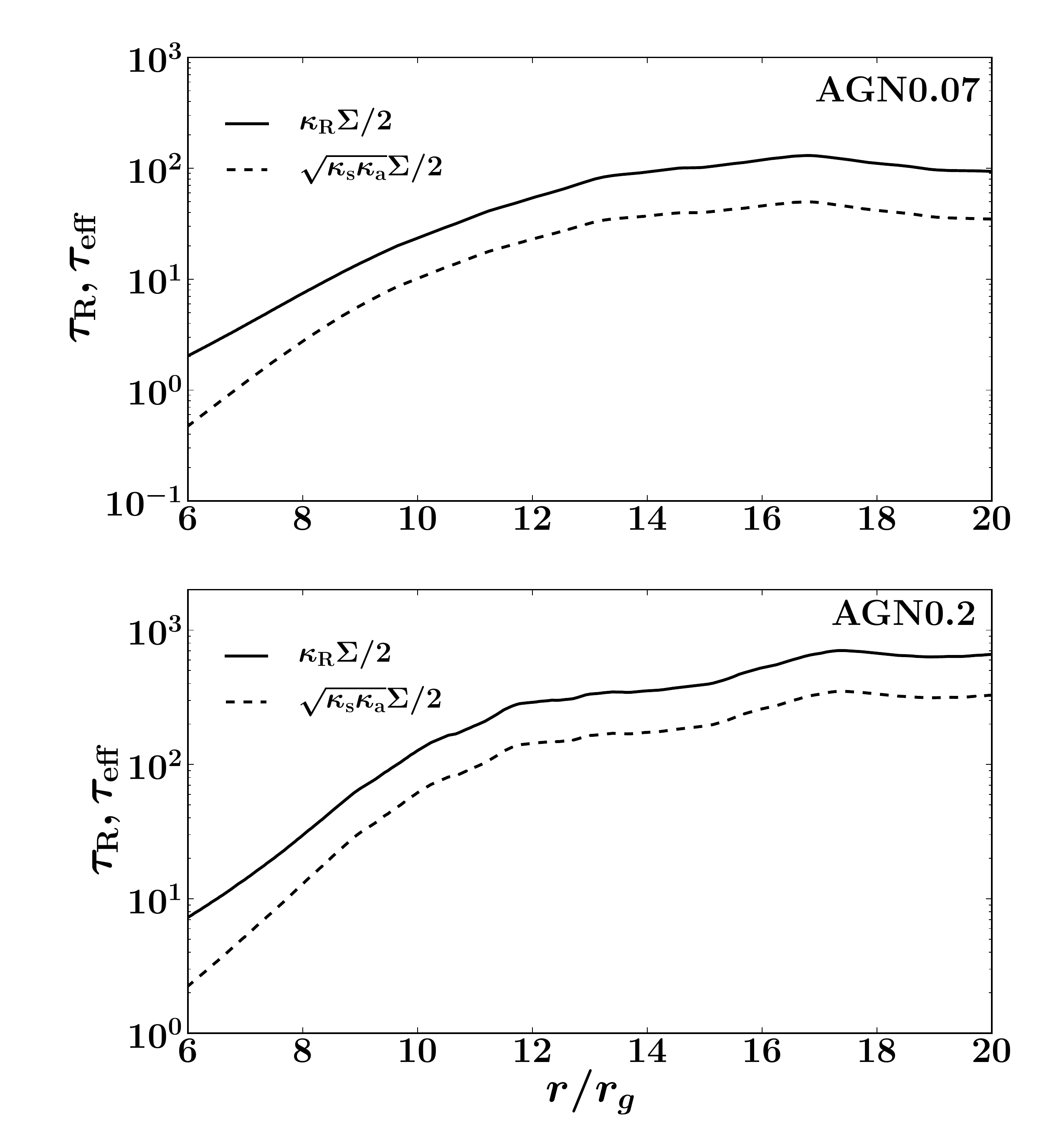}
	\caption{Radial profiles of the total Rosseland mean ($\kappa_R\Sigma/2$) and effective absorption ($\sqrt{\kappa_s\kappa_a}\Sigma/2$) optical depth from the disk midplane to the rotation axis. Here $\kappa_a$ and $\kappa_s$ are the absorption and scattering opacities while $\kappa_R\equiv \kappa_a+\kappa_s$ is the Rosseland mean opacity. The top and bottom panels are for the two runs {\sf AGN0.07} and {\sf AGN0.2} respectively.}
	\label{surface_density}
\end{figure}

\subsection{Disk Structure}
\label{sec:ave_structure}
The time and azimuthally averaged spatial structure inside $30r_g$ for the density, flow velocity, radiation energy density and magnetic field lines of the two simulations are shown in Figure \ref{ave_disk_structure}. Radial profiles of different shell averaged pressure components, the mass accretion rates as well as total optical depth are shown in Figures \ref{pressure_profile},  \ref{Mdot_profile} and \ref{surface_density}. The disk is thinner in the run {\sf AGN0.07} compared with {\sf AGN0.2} due to lower accretion rate and the thickness of the disk is different from standard accretion disk model predictions. 
In a radiation pressure and electron scattering dominated $\alpha$ disk model
in a spherically symmetric gravitational potential,
vertical hydrostatic equilibrium implies that the thickness of the disk
$H=[\kappa_s\dot{M}/(4\pi c)](d\ln\Omega/d\ln r)$ away from the inner boundary.
Near the inner boundary $r_{\rm in}$, it is smaller by $1-\left(r_{\rm in}/r\right)^{1/2}$.
This is nearly independent of radius (and of $\alpha$), and implies $H/r_g\approx 1.1$ \citep{Franketal2002} for $\dot{M}=7\%\Medd$ and $H/r_g\approx 3.0$ for $\dot{M}=20\%\Medd$. It will be even smaller if we take into account the boundary effect. As shown in Figure~\ref{ave_disk_structure}, the heights of the photosphere clearly increase rapidly with radius in the two runs, and become larger than the values in the standard $\alpha$ disk model beyond $\approx 10r_g$. The contrast of radiation energy density inside and above the photosphere is much smaller in {\sf AGN0.07} compared with the run {\sf AGN0.2} because of the reduced total optical depth as well as different spatial distributions of dissipation. Although the shell averaged radiation pressure is still larger or comparable to the magnetic pressure (Figure \ref{pressure_profile}), the vertical gradient of radiation pressure is actually smaller, particularly in the run {\sf AGN0.07}. The disk is actually supported by magnetic pressure gradient in this region (section \ref{sec:vertical}). There are two magnetic field loops above and below the disk midplane in the run {\sf AGN0.07} and they are configured in such a way to have a net radial magnetic field near the midplane. In the run {\sf AGN0.2}, there are net poloidal magnetic fields threaded through the disk by design. 
With both magnetic field configurations, we have confirmed that the strong magnetic pressure is dominated by the turbulent component, since the magnetic pressure due to the azimuthally averaged mean magnetic field $\overline{B}^2/2$ is smaller than the total magnetic pressure by a factor of $\approx 10$ (Figure \ref{pressure_profile}).

Radial profiles of mass accretion rate averaged over the whole spherical shell reach constant values inside $\approx 15r_g$ for {\sf AGN0.07} and $\approx 13r_g$ for {\sf AGN0.2} (Figure \ref{Mdot_profile}). We decompose the net mass accretion rate into two parts based on the location of the photosphere shown in Figure \ref{ave_disk_structure}. At each radius, we integrate $\rho v_r r^2$ for $\theta$ angles either above or below the photosphere and radial profiles of the two components are shown as red and dashed black lines in Figure \ref{Mdot_profile}. For the run {\sf AGN0.2}, the majority of accretion happens in the optically thick part of the disk until $r<7r_g$. While for the run {\sf AGN0.07}, more than half of the mass is accreted in the optically thin surface of the disk
over the entire radial range where the disk has reached steady state. The spatial distribution of mass inflow is consistent with the location of stress for angular momentum transport (section \ref{sec:stress}). Significant surface accretion has also been found in non-radiative ideal MHD simulations of accretion disks with a net poloidal magnetic field \citep{ZhuStone2018}. In fact, the resulting magnetic field configuration from the run {\sf AGN0.2} shown in Figure \ref{ave_disk_structure} is very similar to what \cite{ZhuStone2018} found (for example, Figure 25 of that paper). The poloidal magnetic fields are advected inwards above the disk midplane. Since we can determine the thermal properties of the disk self-consistently, for the accretion rate of $26\%\Medd$, accretion still happens in the optically thick part of the disk with little accretion in the corona region. When the accretion rate drops to $7\%$ with a reduced optical depth, mass is primarily accreted in the optically thin region. However, for these AGN disks, the mechanism for angular momentum transport responsible for the corona accretion is due to radiation instead of magnetic field as found in non-radiative ideal MHD simulations (section \ref{sec:stress}).

Radial profiles of time averaged Rosseland mean optical depth $\tau_R=\kappa_R\Sigma/2$ for the two runs are shown in Figure \ref{surface_density}. Here the Rosseland mean opacity $\kappa_R\equiv \kappa_a+\kappa_s$ is the sum of absorption $\kappa_a$ and scattering $\kappa_s$ opacity. Optical depths for effective absorption $\tau_{\rm eff}=\sqrt{\kappa_s\kappa_a}\Sigma/2$ are also shown in the same Figure. The Rosseland mean optical depth is only 2 at $6r_g$ for {\sf AGN0.07} and it increases to $100$ at $14r_g$, while the effective absorption optical depth drops below $1$ near the ISCO. When the accretion rate is increased to $26\%\Medd$, the optical depth is increased by a factor of $2-3$ and the entire disk is optically thick even for effective absorption. The optical depth in the disk is smaller than what the standard $\alpha$ disk model predicts by a factor of $\approx 10$ for the same mass accretion rate due to increased disk scale height and inflow velocity because of strong magnetic pressure support (Section \ref{sec:vertical}). 
%The fact that the inner region of the disk can become effective optically thin suggests that this %part of \citep{Halletal2018} the change of disk spectrum (non-blackbody)

\begin{figure}[htp]
	\centering
	\includegraphics[width=1.0\hsize]{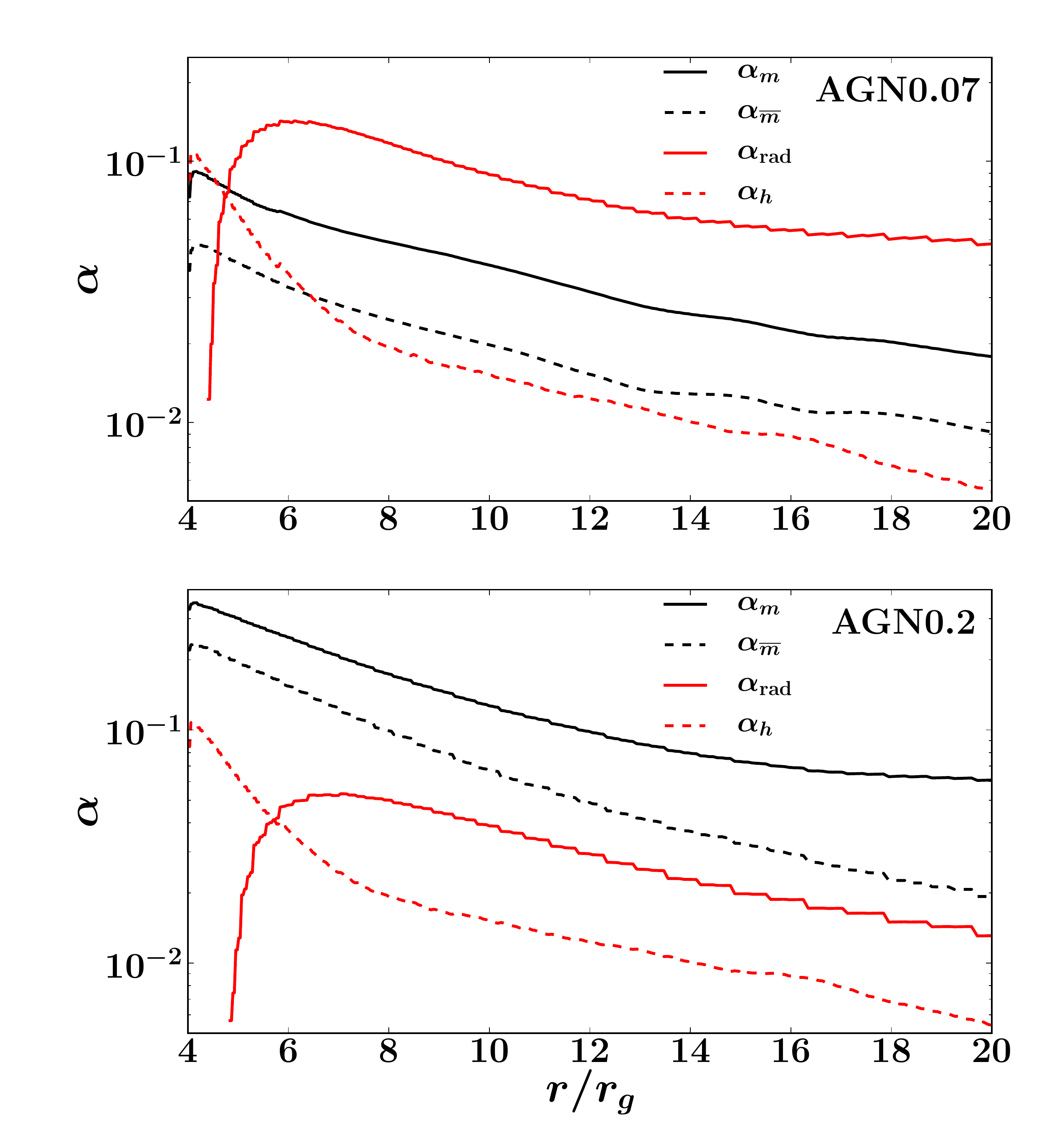}
	\caption{Radial profiles of the time and shell averaged ratios between Reynolds ($\alpha_h$, dashed red lines), Maxwell ($\alpha_m$, solid black lines), radiation ($\alpha_{\rm rad}$, solid red lines) stresses and total pressure for the run {\sf AGN0.07} (top) and {\sf AGN0.2} (bottom). The dotted black lines ($\alpha_{\overline{m}}$) are for stresses due to the azimuthally averaged mean $\langle B_r\rangle$ and mean $\langle B_{\phi}\rangle$.}
	\label{compare_alpha}
\end{figure}

\begin{figure*}[htp]
	\centering
	\includegraphics[width=1.0\hsize]{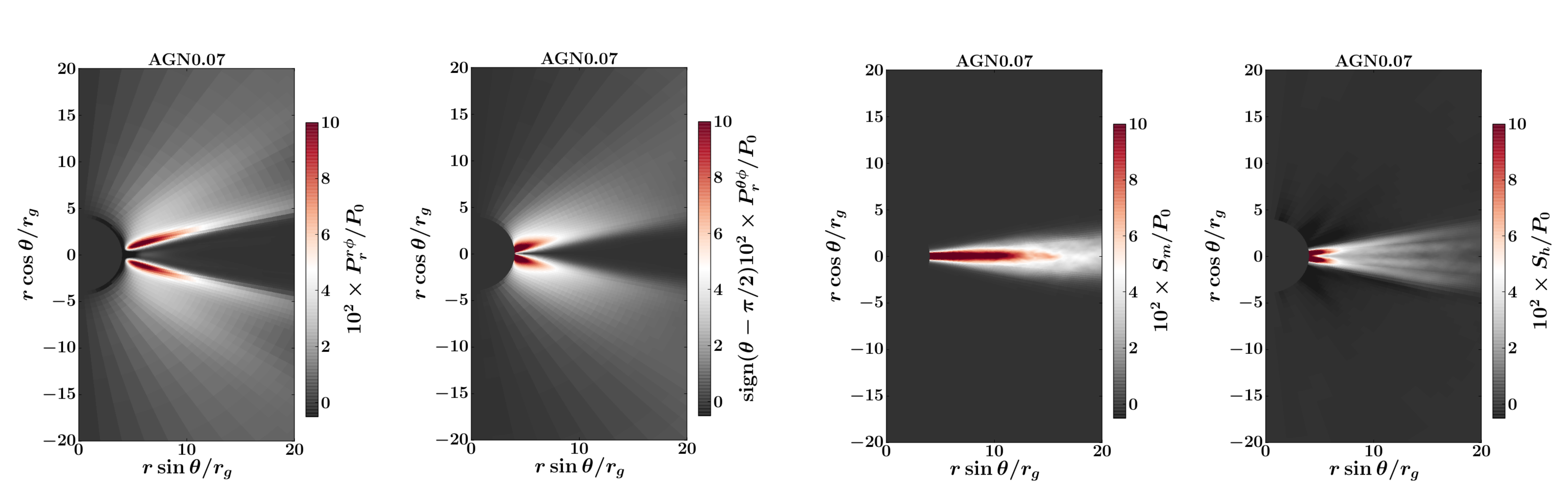}
	\caption{Time and azimuthally averaged spatial distributions for the off-diagonal components of radiation pressure $P_r^{r\phi}$ (the first panel) and $P_r^{\theta\phi}$ (the second panel) for the run {\sf AGN0.07}. We add a negative sign to $P_r^{\theta\phi}$ above the disk midplane so that it can be compared with other stress component easily with the same color scheme. The third and fourth panels are the corresponding Maxwell and Reynolds stresses for comparison.}
	\label{AGN3Bstress}
\end{figure*}

\subsection{Stresses for Angular Momentum Transport}
\label{sec:stress}
In accretion disks, transport of angular momentum can be provided by the sum of Maxwell, Reynolds and radiation (the off-diagonal components of the radiation pressure) stresses. Since radiation pressure is so significant in AGN accretion disks, radiation stress can potentially play an important role as an effective viscosity (\citealt{LoebLaor1992}). In the optically thick regime, radiation viscosity is proportional to $E_r/(c\rho\kappa_R)$ multiplied by the shear rate of the velocity field (\citealt{Weinberg1971,MihalasMihalas1984,KaufmanBlaes2016}). If we only consider the shear rate for a Keplerian disk near the disk midplane, the ratio between the radiation stress and radiation pressure at radius $r$ is $\sim \mathcal{O}\left(\sqrt{r_g/r}/(r\rho\kappa_R)\right)$. For a standard thin accretion disk model where $r\rho\kappa_R \gtrapprox 100$, the radiation stress will be much smaller than the typical Maxwell and Reynolds stresses produced by MRI turbulence. In other words, radiation viscosity is small because the mean free path of the photons is too short. This is also true for super-Eddington accretion disks around AGNs due to large optical depth \citep{Jiangetal2018}, even though radiation pressure is significantly larger than the gas and magnetic pressure there.
However, this is not the case for these sub-Eddington simulations.

\begin{figure}[htp]
	\centering
	\includegraphics[width=1.0\hsize]{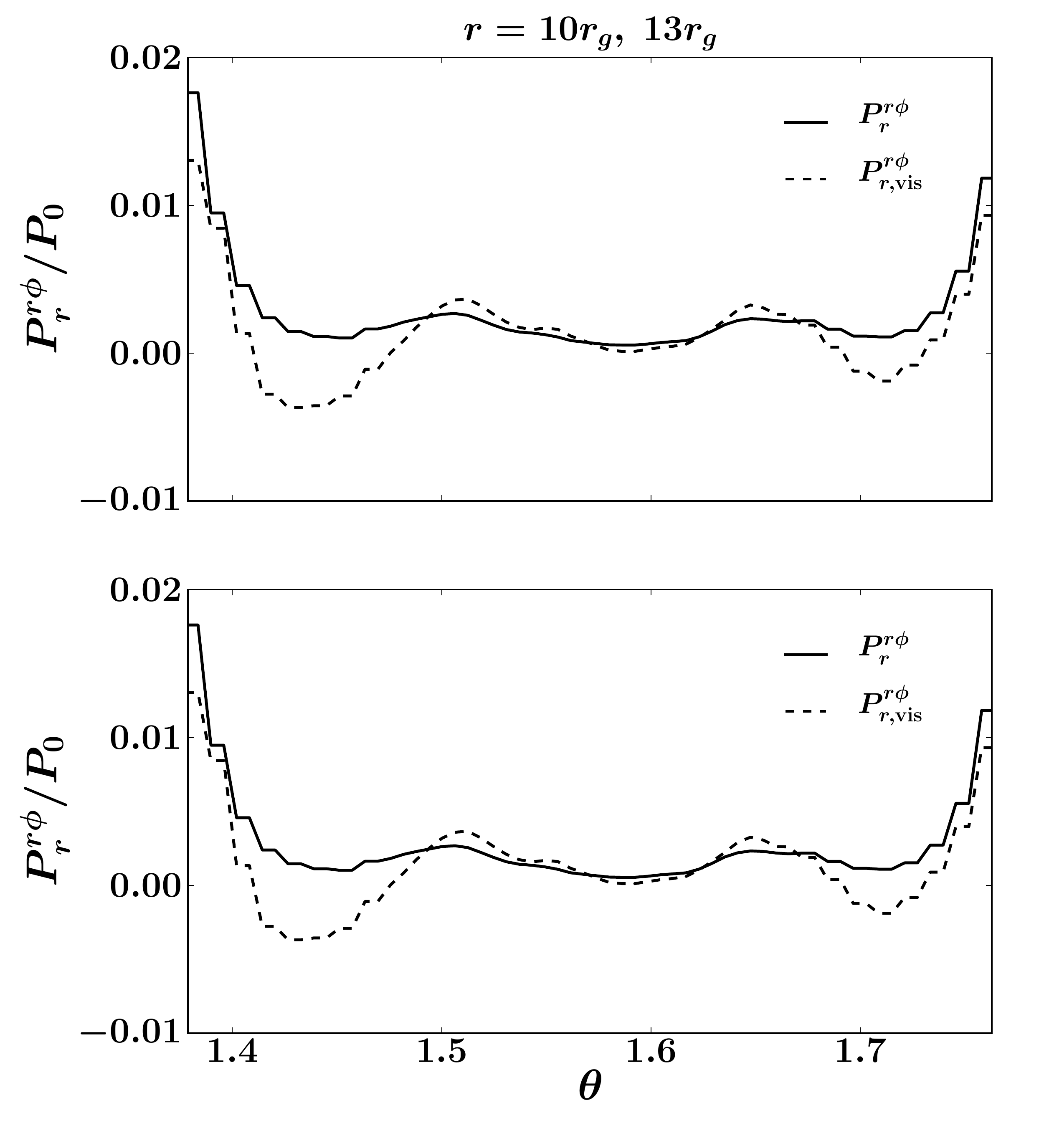}
	\caption{Comparison between the vertical profiles of the azimuthally
averaged radiation stress $P_r^{r\phi}$ (solid lines) calculated by the simulation {\sf AGN0.07} with the analytical formula for radiation viscosity $P_{r,\rm vis}^{r\phi}$ (dashed lines), based on equations (\ref{eq:vis1}) and (\ref{eq:vis2}). The top and bottom panels are for radii $10r_g$ and $13r_g$ respectively. This comparison is done for the snapshot at time $4.5\times 10^4r_g/c$.}
	\label{compare_vis_stress}
	\vspace{0.1cm}
\end{figure}

\begin{figure}[htp]
	\centering
	\includegraphics[width=0.49\hsize]{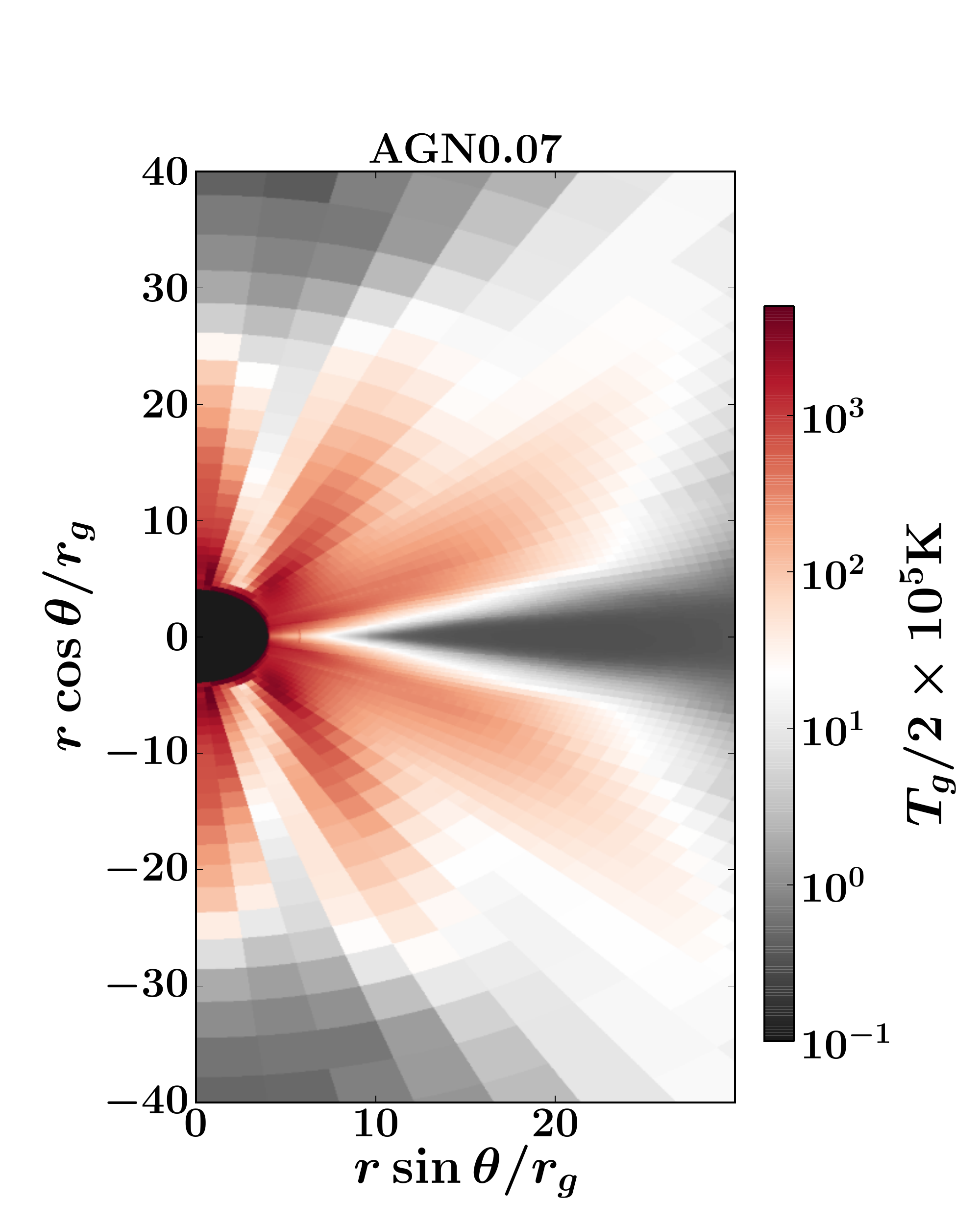}
	\includegraphics[width=0.49\hsize]{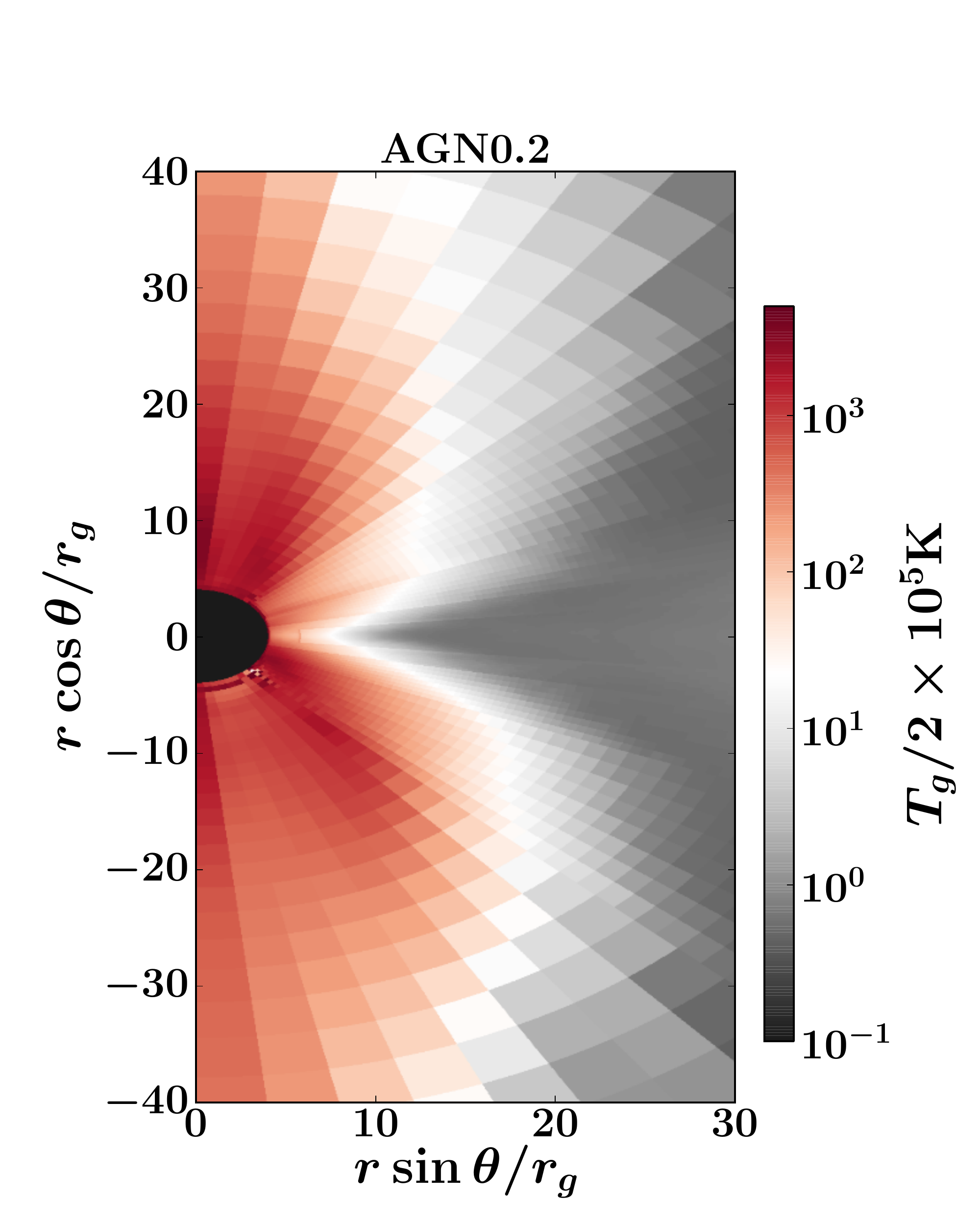}
	\caption{Time and azimuthally averaged spatial distributions of gas temperature $T_g$ for the two runs {\sf AGN0.07} (left panel) and {\sf AGN0.2} (right panel). The gas temperature is $\approx 10^5-2\times 10^5$ K in the optically thick part of the disk but increases to $10^8-10^9$ K rapidly in the optically thin corona regions.  }
	\label{Corona}
\end{figure}

\begin{figure*}[htp]
	\centering
	\includegraphics[width=0.49\hsize]{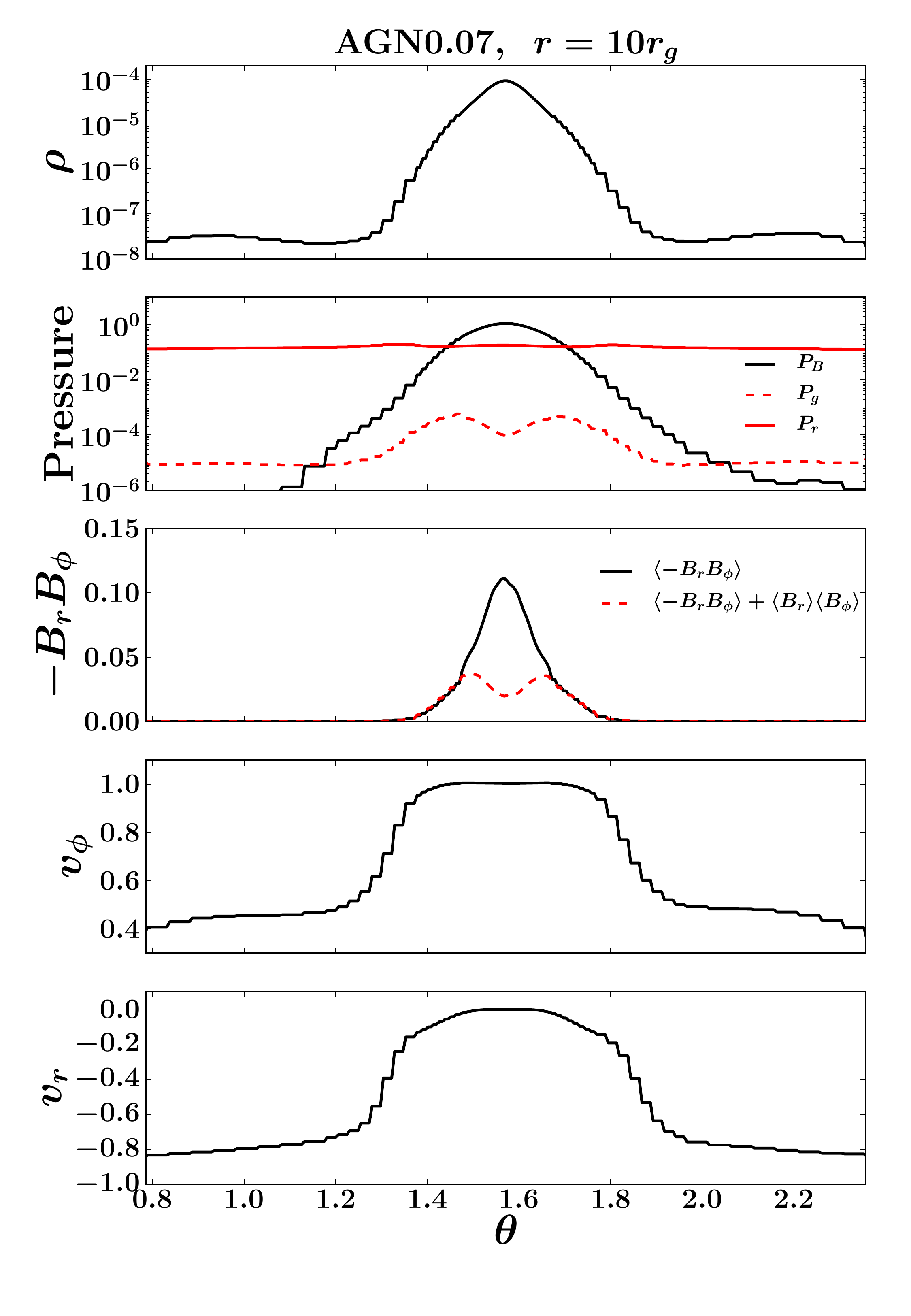}
	\includegraphics[width=0.49\hsize]{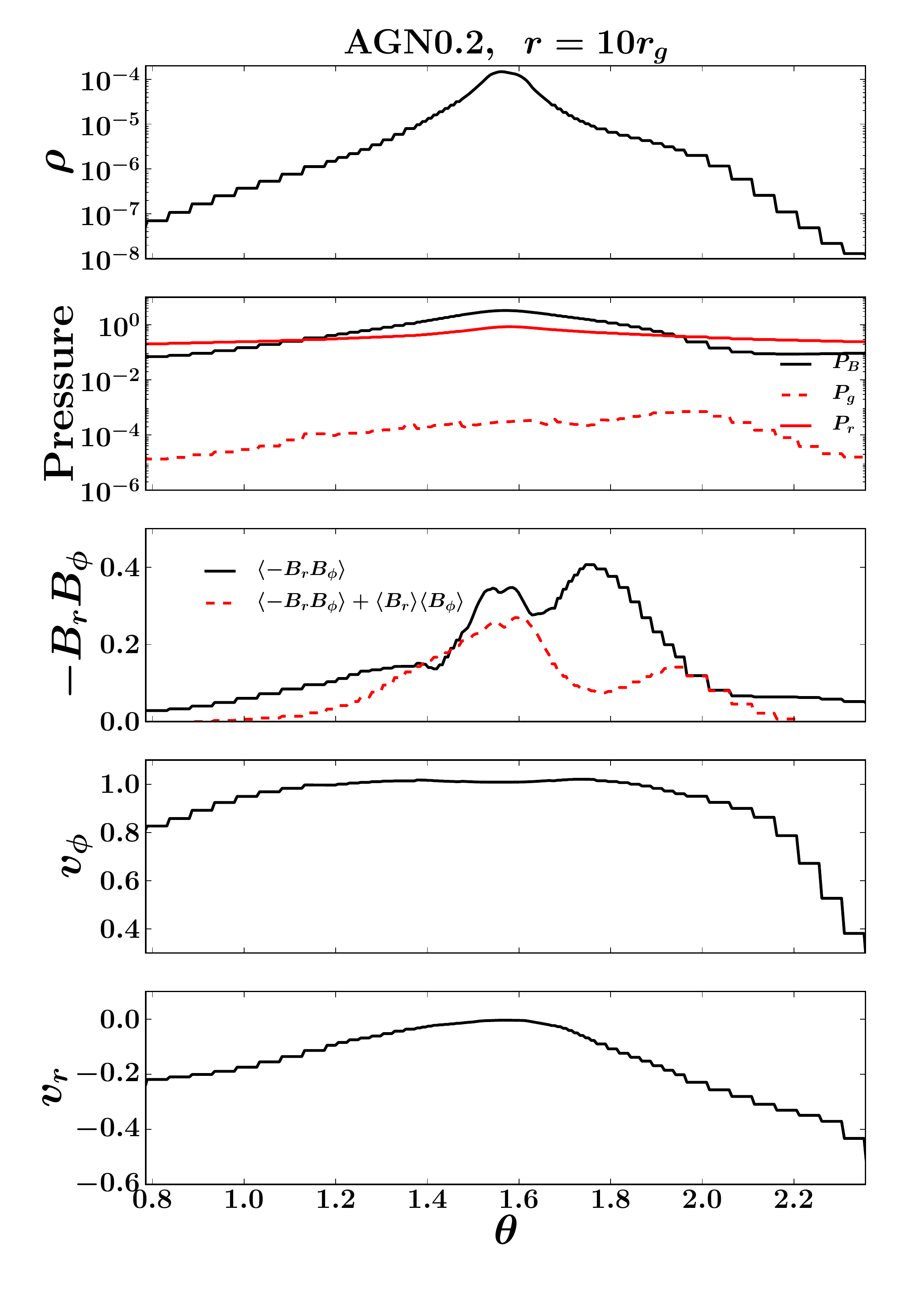}
	\caption{Time and azimuthally averaged poloidal profiles of various quantities at $10r_g$ for the run {\sf AGN0.07} (left panels) and {\sf AGN0.2} (right panels). From top to bottom, these quantities are density $\rho$ at the top panel, gas ($P_g$, dashed red lines), radiation ($P_r$, solid red lines) and magnetic ($P_B$, solid  black lines) pressure in the second panel, total Maxwell stress (solid black lines) as well as the turbulent component (dashed red lines) in the third panel, density weighted rotation (fourth panel) and radial velocities (bottom panel). Density is in unit of $\rho_0$ while pressure and Maxwell stress are in unit of $P_0$. The velocities are scaled with the Keplerian value at the midplane $\sqrt{rG\mbh}/(r-2r_g)$. }
	\label{vertical}
\end{figure*}

\begin{figure}[htp]
	\centering
	\includegraphics[width=1.0\hsize]{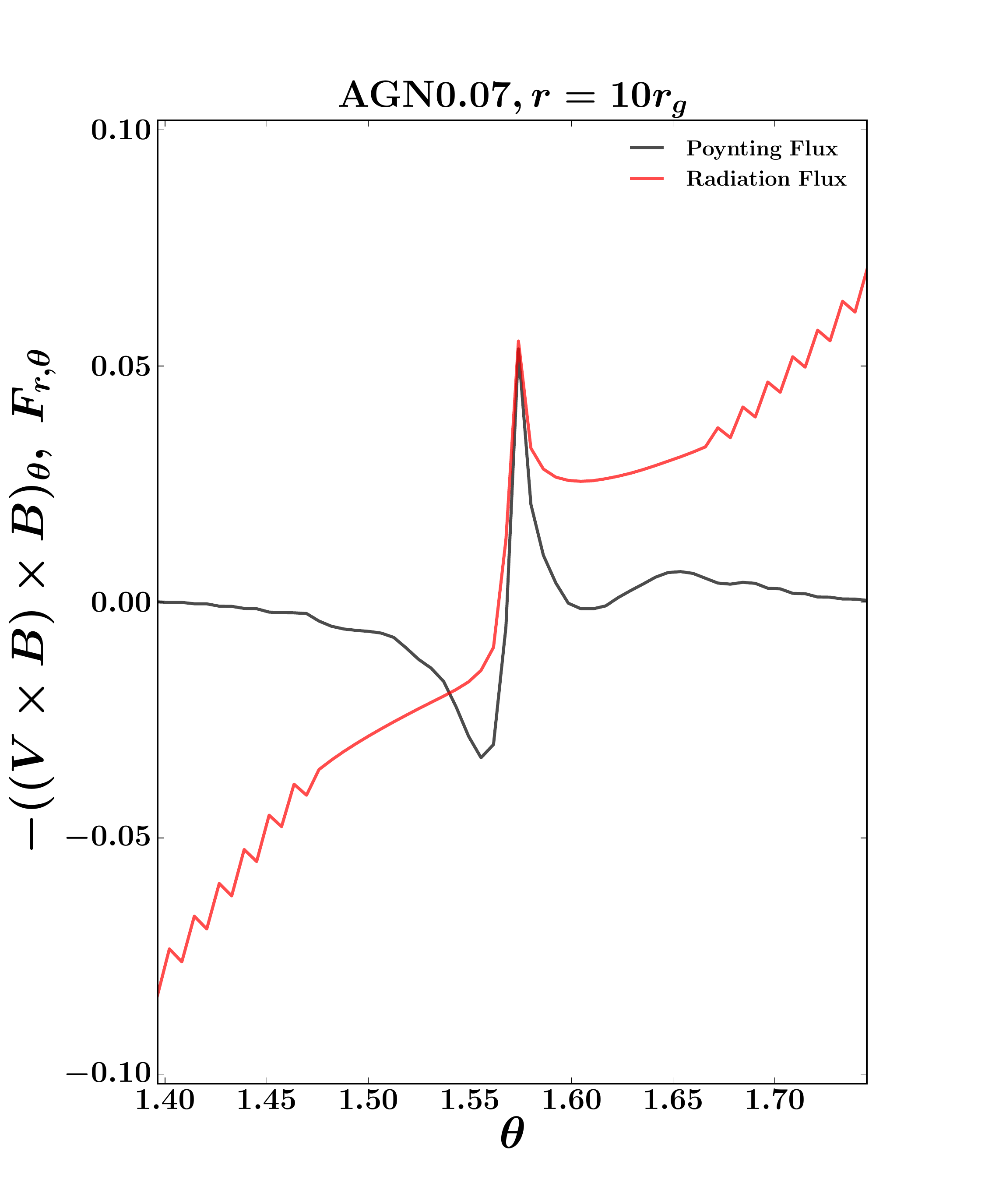}
	\caption{Time and azimuthally averaged vertical profiles of the poloidal components of Poynting flux (black line) and radiation flux (red line) at $10r_g$ for the run {\sf AGN0.07}. The energy fluxes are scaled with the critical value $c G\mbh|\cos(\theta)|/\left(\kappa_{\rm es}(r-2r_g)^2\right)$.}
	\label{vertical_energy}
\end{figure}

Radial profiles of the shell averaged Maxwell ($\alpha_m$), Reynolds ($\alpha_h$) and radiation stresses (only the $r-\phi$ component remains, $\alpha_{\rm rad}$) scaled with the shell averaged total pressure for the two runs are shown in Figure \ref{compare_alpha}. The stresses are calculated in the same way as in \cite{Jiangetal2018}. For both runs, we have also calculated the Maxwell stress due to the azimuthally averaged mean $\langle B_r \rangle$ and $\langle B_{\phi} \rangle$, which are shown as the dashed red lines in Figure \ref{compare_alpha}. It is smaller than the total Maxwell stress by a factor of 6 in {\sf AGN0.2} and a factor of 2 in {\sf AGN0.07}, despite the fact that the latter has shearing net radial magnetic field in the midplane. The total stress in the run {\sf AGN0.2} is dominated by the Maxwell stress with the Reynolds and radiation stresses smaller by a factor of $\approx 5$, which is consistent with the results reported by previous both local and global simulations of MRI turbulence with net vertical magnetic fields \citep{BaiStone2013,ZhuStone2018}. However, for the run {\sf AGN0.07}, the radiation stress is larger than the others by more than a factor of $2$ for all the radial range $r>5r_g$, while the Maxwell and Reynolds stresses vary from $2\%$ to $10\%$ of the total pressure, which are typical values found by MRI turbulence without net vertical magnetic fields \citep{HGB1995,MillerStone2000,Hiroseetal2006,Davisetal2010,Jiangetal2013c,Jiangetal2014}.

Radiation stress and Maxwell stress dominate different locations of the disk in  the run {\sf AGN0.07} as they have different spatial distributions. 
The time and azimuthally averaged spatial distributions of the $r-\phi$ and $\theta-\phi$ components of radiation stress $P_{r}^{r\phi}$ and $P_{r}^{\theta\phi}$ for the run {\sf AGN0.07} are shown in the first two panels of Figure \ref{AGN3Bstress}. For comparison, the Maxwell stress $S_m$ and Reynolds stress $S_h$ are shown in the third and fourth panels. Inside the photosphere of the disk, the radiation stress is actually much smaller than the Maxwell and Reynolds stresses. This is consistent with our estimate that radiation stress is proportional to $1/(r\rho\kappa_R)$, while the Maxwell and Reynolds stresses roughly follow the distribution of mass. However, once near the photosphere, because of the increased photon mean free path as well the large velocity gradient,  significant radiation stress shows up while Maxwell and Reynolds stresses drop significantly. 
This suggests that in the run {\sf AGN0.07}, angular momentum is transported by the Maxwell and Reynolds stresses in the optically thick part of the disk. But once near the photosphere, angular momentum is transported by the radiation stress. Since total radiation stress dominates, more accretion happens in the optically thin region, which is consistent with the mass accretion rates shown in Figure \ref{Mdot_profile}. This is not the case for the run {\sf AGN0.2}, where most accretion still happens in the optically thick region until $r \lessapprox 8r_g$. This also causes a stronger corona component at large radii for the run {\sf AGN0.07} (Section \ref{sec:corona}), since dissipation associated with the accretion in the optically thin region can heat up the gas easily without cooling efficiently.

%Since our simulations have net radial magnetic field near the disk midplane, which will be sheared %to form $B_{\phi}$ and result in Maxwell stress even without the turbulence. This component can be %quantified by calculating the azimuthally averaged $\langle B_r\rangle$ and $\langle B_{\phi} %\rangle$ and the associated stress $-\langle B_r\rangle \langle B_{\phi} \rangle$. Radial profiles %of the ratio between the shell averaged stress from this component and total pressure %$\alpha_{\overline{m}}$ are shown as the dotted red lines in Figure \ref{compare_alpha}. This %component is smaller than the total Maxwell stress by a factor of $\approx 2$ in {\sf AGN0.07} and a %factor of $\approx 8$ in {\sf AGN0.2}. This demonstrates that the dominant Maxwell stress is still %from the turbulent component generated by MRI.

The amount of radiation stress we observe in the simulations can also be
compared with analytic expectations for radiation viscosity. When the
optical depth per cell is large, the $r-\phi$ component of the co-moving
frame radiation viscosity can be calculated analytically as
\citep{Masaki1971,KaufmanBlaes2016}
\begin{eqnarray}
P_{r,{\rm vis}}^{r\phi}=-\frac{8}{27}\frac{E_r}{\rho\kappa_R c}D_{r\phi},
\label{eq:vis1}
\end{eqnarray}
where the $r-\phi$ shear rate in spherical polar coordinates is
$D_{r\phi}=r\partial \left(v_{\phi}/r\right)/\partial r$+$1/\left(r\sin\theta\right)\partial v_r/\partial \phi$. However, when cells become optically thin,
this formula needs to be modified to account for the nonzero mean free path of photons.  Under optically thin conditions, the viscous stress tensor associated with a shear velocity difference across some given length scale should be proportional to the optical depth across that scale times the velocity difference \citep{Socratesetal2004,KaufmanBlaes2016}.
In order to ensure continuous behavior
between the optically thin and thick regimes, we therefore write the
stress tensor as
\begin{eqnarray}
P_{r,{\rm vis}}^{r\phi}=-\frac{8E_r}{27c}k(\tau)D^{\prime}_{r\phi},
\label{eq:vis2}
\end{eqnarray}
Here $D^{\prime}_{r\phi}\equiv r\Delta (v_{\phi}/r)+\Delta v_r$ is just the difference of shear rate across distances
with optical depth $\tau$ along radial (for $v_{\phi}$) and azimuthal (for $v_r$) directions. 
 and $k(\tau)=\tau^{-1}$ for $\tau>1$ and $\tau$ for $\tau<1$. In practice, we first determine the optical depth per cell at each location. If that value is larger than one, we calculate the difference of shear rate between neighboring cells. If that value is smaller than 1, we extend the distance until $\tau=1$ is reached and calculate the difference of shear rate there. 
The radiation stress tensor calculated based on the above formula is transformed back to the lab frame and then averaged azimuthally to compare with the $r-\phi$ radiation stress returned by the simulations directly. Comparisons for two different radii are shown in Figure \ref{compare_vis_stress}. The numerically calculated radiation stress nicely follow the prediction for radiation viscosity in both optically thick and optically thin regimes.

\subsection{Vertical Structure of the Disk}
\label{sec:vertical}
Time and azimuthally averaged poloidal profiles of various quantities at $10r_g$ for the two runs are shown in Figure \ref{vertical}. Density drops faster with height in the run {\sf AGN0.07} due to a smaller pressure scale height. Shapes of density profiles are also more centrally peaked compared with gas or radiation pressure dominated disks found by both local shearing box and global simulations \citep{Turner2004,Hiroseetal2006,Hiroseetal2009,Jiangetal2016a,Jiangetal2018}. 
Although the shell averaged radiation pressure is comparable to the shell averaged magnetic pressure in the run {\sf AGN0.07} (Figure \ref{pressure_profile}), the radiation pressure is relatively flat with height and the whole disk is supported by the magnetic pressure gradient. In fact there is a small enhancement of $P_r$ near the photosphere (see also Figure \ref{ave_disk_structure}) due to the significant dissipation there caused by radiation viscosity.
In the run {\sf AGN0.2}, radiation pressure partially supports the disk near the midplane and the disk becomes completely magnetic pressure supported around $10^{\circ}$ away from the midplane. This also causes the density to drop more slowly with height around the same location as shown in the right panel of Figure \ref{vertical}. Gas pressure is always smaller than the other pressure components by more than a factor of 1000. 

For the two magnetic field configurations used in the simulations, besides the Maxwell stress from the turbulence, there are always significant azimuthally averaged mean $\langle B_r\rangle$ and $\langle B_{\phi}\rangle$, although the product of these components never becomes the dominant stress (Figure \ref{compare_alpha}). This is different from the magnetic pressure supported disk as found by \cite{Gaburovetal2012}, where the Maxwell stress is primarily due to $-\langle B_r\rangle \langle B_{\phi}\rangle$.
They also have different vertical distributions compared with the turbulent stress as shown in the third panels of Figure \ref{vertical}.  The Maxwell stress generated by the turbulence (the dashed red lines) shows double peaks away from the midplane, which is consistent with previous simulations \citep{Blaesetal2011,Jiangetal2016a}. The stress due to the azimuthally averaged mean magnetic field (the difference between the solid black lines and the dashed red lines) is peaked at the midplane and drops quickly with height.

Inside the photosphere near the midplane, the rotation speed of the disk is pretty close to the Keplerian value with negligible radial inflow speed (the fourth and bottom panels of Figure \ref{vertical}). Inside the corona region, the rotation speed drops to only $40\%$ of the Keplerian value and significant radial velocity is present. 

Despite the fact that the disk is supported by magnetic pressure, the energy dissipated in the disk is transported away vertically by both the radiation flux and Poynting flux near the midplane. Vertical profiles of the poloidal components of the energy fluxes at $10r_g$ for the run {\sf AGN0.07} are shown in Figure \ref{vertical_energy}. They are scaled with the critical energy flux $F_c\equiv cG\mbh|\cos(\theta)|/\left(\kappa_{\rm es}(r-2r_g)^2\right)$, which is the radiation flux if the vertical component of gravity is completely balanced by the radiation force with the electron scattering opacity $\kappa_{\rm es}$. The radiation flux is less than $\approx 5\%$ of the critical value inside the disk, which confirms that radiation pressure plays a negligible role to support the disk. But the radiation flux is comparable to the Poynting flux within $\approx 3^{\circ}$ away from the midplane and it completely dominates the energy transport beyond that. The sign of Poynting flux also suggests that magnetic energy amplified in the disk is transported away from the midplane likely due to magnetic buoyancy \citep{Blaesetal2011,Begelmanetal2015}, although this simulation does not show any butterfly diagram (Figure \ref{STplot}). The existence of non-zero azimuthally averaged radial magnetic field $\langle B_r\rangle$ in localized region of the disk will also keep generating $B_{\phi}$ due to the differential rotation of the disk, which is balanced by the buoyant escape of the field. Since the shell averaged $B_r$ is zero, $B_{\phi}$ generated at different locations in the disk will reconnect and cause dissipation.

\subsection{Properties of the Coronae}
\label{sec:corona}
Local shearing box radiation MHD simulations \citep{Jiangetal2014} have found that coronae with high gas temperature are only formed above the disk when the surface density is low enough so that a significant fraction of the total dissipation happens in optically thin regions.  For the surface density at $30r_g$ adopted by \cite{Jiangetal2014} based on the standard thin disk model, only $3.4\%$ of the dissipation happens in the corona region. This fraction is increased to more than $50\%$ inside $15r_g$ for the run {\sf AGN0.07} with significantly reduced surface density in the magnetic pressure supported disk. This results in a high gas temperature of $\approx 10^8$ K covering all the radial range $\lesssim 15r_g$ with height $20r_g$ away from the disk midplane as shown in the left panel of Figure \ref{Corona}. When the accretion rate is increased to $26\%\Medd$ in the run {\sf AGN0.2}, the disk becomes thicker and the fraction of accretion as well as the associated dissipation in the optically thin region are also reduced. Coronae with high gas temperature ($\gtrapprox 10^8$ K) only show up inside $\lesssim 10r_g$ when the surface density drops to a value such that $\kappa_R\Sigma/2\lesssim 100$.  Inside this region, the corona is also more vertically extended covering more than $40r_g$ away from the disk midplane. 

Coronae produced by these simulations are pretty consistent with many observational inferred properties of AGN corona. The simulations only produce high temperature corona (gas temperature $>10^8$ K) inside $\approx 10r_g$ and the temperature decreases with increasing distance from the black hole. This is because surface density of the disk increases with radius and the fraction of dissipation in the optically thin region drops, which is necessary to maintain the high temperature corona. This agrees with the constrains from both reflection and microlensing modeling of AGN coronae \citep{Chartasetal2009,Daietal2010,ReisMiller2013,JimenezVicenteetal2015}, which suggest that they are compact with roughly the same size as we find. 

Surface density increases for the same region of the disk when the accretion rate increases from {\sf AGN0.07} to {\sf AGN0.2}. This is in contrast to the standard thin disk model, for which the surface density decreases linearly with increasing mass accretion rate in the radiation pressure dominated regime. When the surface density increases, the fraction of dissipation in the optically thin regime is reduced and the radial regime where the corona is formed is also smaller. This will result in a smaller ratio between X-ray luminosity from the corona and luminosity of the thermal emission from the disk, which is in excellent agreement with the observational fact that the spectral index of AGNs gets harder as luminosity decreases \citep{Steffenetal2006,Justetal2007,Ruanetal2019}.

\section{Discussion}
\label{sec:discussion}
It is interesting to compare the structures of the disk produced by the simulations with proposed models of magnetic pressure dominated disks in the literature. \cite{Parievetal2003} modified the standard $\alpha$ disk model by including a magnetic pressure component in the disk, which was assumed to be larger than the thermal pressure at all radii. Magnetic pressure is assumed to be a constant at each radius and decreases with increasing radius following a power law, the slope of which is a free parameter. The strength of the magnetic field is also a free parameter in this model. The disk structures in this model are very similar to the original $\alpha$ disk model, which predicts that the surface density is proportional to $1/\dot{M}$. This is clearly in contrast to what we find from the simulations. \cite{BegelmanPringle2007} proposed that the toroidal magnetic field  should saturate to the level so that the associated Alfv\'en velocity is the geometric mean of gas isothermal sound speed $c_s$ and Keplerian speed $V_k$ . Under this assumption, the disk surface density is found to be proportional to $ \dot{M}^{7/9}$, which has the same trend as what we find, although their dependence on $\dot{M}$ is weaker. The Alfv\'en velocity in our simulations is found to be larger than $\sqrt{c_sV_k}$ by a factor of $2-3$ in the midplane and much smaller than that near and above the photosphere. Our simulation results can be used to constrain the assumptions and improve these analytical models.

The relatively uniform spatial distribution of $E_r$ in the run {\sf AGN0.07} is caused by the significant dissipation near the photosphere. In fact, $E_r$ near the photosphere is slightly larger than $E_r$ at the midplane as shown in the left panel of Figure \ref{vertical}. This is also consistent with the rapid increase of radiation flux at these locations as shown in Figure \ref{vertical_energy}. For comparison, when more dissipation happens in the optically thick part of the disk as in the run {\sf AGN0.2}, radiation energy density decreases with increasing height monotonically. This is also the reason why radiation viscosity can be significant in the run {\sf AGN0.07}. Since radiation viscosity is proportional to $E_r/\tau$ in the optically thick regime, if $E_r$ at the photosphere is smaller than $E_r$ in the midplane by a factor of $1/\tau$ as in the standard thin disk model, radiation viscosity will not be significant near the photosphere even mean free path is larger there.

These simulations adopt the gray approximation for radiation transport and Compton scattering is included based on the effective radiation temperature to roughly estimate 
the energy exchange between photons and gas. Therefore, we cannot produce the expected spectra from these simulations directly. This approximation will also cause 
underestimate of gas temperature in the corona region. 
Monte Carlo technique has been developed to post process the simulation data and generate spectra from the disk. Preliminary results show that the coronae in these simulations are indeed able to produce significant X-ray emission with X-ray luminosity consistent with the total dissipation we have in these region. Detailed spectrum properties will be presented in a separate publication. Since corona only exists inside $\approx 10r_g$, for which general relativity may play an important role, simulations with fully self-consistent radiation transport in general relativity will be carried out in the near future.

The total luminosity from the inner $10r_g$ of the disk is only $1\%\Ledd$ with $\Medd=7\%\Medd$ for the run {\sf AGN0.07}, which means the radiative efficiency is only $1.4\%$. The radiative efficiency is slightly increased to $\approx 3.5\%$ for the run {\sf AGN0.07}. The efficiency from these disks is significantly smaller than the prediction of the standard thin disks for the same radial range. One reason for the lower efficiency is the significant dissipation in the optically thin region, which cannot convert to thermal radiation efficiently. Another reason for the lower radiative efficiency in the run {\sf AGN0.07} is the stress  $-\< B_r\>\<B_{\phi}\>$ near the disk midplane (Figure \ref{vertical}). This component of stress can cause mass accretion without dissipation.

%Because both the disk and corona are simulated under the MHD equation and we do not distinguish the electron %and proton temperatures, we will very unlikely get the very high energy part of the X-ray spectrum. 

Both of the simulations in this paper established long-lived thermal equilibria
between turbulent dissipation of gravitational binding energy in the flow and
radiative cooling for more than $10$ thermal time scales.  This is despite the fact that radiation dominates the
thermal pressure, a situation which generally leads to thermal instability
in $\alpha$-disk models \citep{ShakuraSunyaev1976}, at least when electron
scattering dominates the Rosseland mean opacity.  Iron opacity can be
significant in disks around supermassive black holes, and has been shown
to stabilize the thermal instability in shearing box simulations of MRI
turbulence \citep{Jiangetal2016a}.  However, while we have included this
source of opacity in the simulations here, it actually does not play a
significant role in the inner regions near the black hole.  Instead, it is
likely that the fact that these simulations are supported by magnetic, not
thermal pressure, leads to their stability \citep{BegelmanPringle2007}.  Indeed,
\citet{Sadowski2016} found that global simulations of geometrically thin
disks are stabilized if the magnetic pressure exceeds half the total pressure,
a criterion which we more than satisfy in our simulations here.

Light curves of a variety of accretion-powered sources in astrophysics, including active galactic nuclei, show a linear relationship between rms variability amplitude and flux (see e.g. \citealt{Uttleyetal2005} and references therein).  Thereis a hint of this in the {\sf AGN0.2} light curve shown in Figure~\ref{mdot_hist}, where the variability amplitude is clearly larger at higher luminosities than at lower luminosities.  Unfortunately, the simulation was not run long enough and the light curve is too short to produce a clean rms-flux relation.  On the other hand, there is no indication of any relationship between variability amplitude and flux in {\sf AGN0.07}.  \citet{HoggReynolds2016} have suggested that the rms-flux relation is related to the presence of the characteristic butterfly diagram of quasi-periodic azimuthal field reversals.  It is therefore noteworthy that these are present only in {\sf AGN0.2}, and not in {\sf AGN0.07} because of its much stonger magnetic pressure support.  The presence of an rms-flux relation may be a way of observationally ruling out magnetic fields which are so strong that the butterfly dynamo is suppressed, but longer simulations will need to be run to confirm this.

Note that the shape and magnetic field in the initial tori we use for the two simulations {\sf AGN0.07} and {\sf AGN0.2} are very similar to the ones adopted by \cite{Jiangetal2017b}. We only change the density of the torus to achieve different mass accretion rates, while the ratio between the initial magnetic pressure and radiation pressure is kept fixed. However, the accretion disks with super-Eddington accretion rates shown in \cite{Jiangetal2017b} do not evolve to a magnetic pressure dominated regime. When the surface density as well as the optical depth are increased, radiation pressure becomes more important in supporting the disk. Since the amount of magnetic field that in the disk is determined by a balance between buoyant escape and amplification by the MRI, this suggests that stronger radiation support in the disk may either increase the magnetic buoyancy or reduce the saturation amplitude of MRI \citep{Jiangetal2013b}. We intend to investigate this further in future work.  

\section{Summary}
We have successfully simulated two accretion disks around a $5\times 10^8\mbh$ black hole with mass accretion rates reaching $0.07\Medd$ and $0.2\Medd$ up to $\sim 15r_g$. The disks do not show any sign of thermal instability over many thermal timescales. The structure of the disk differs markedly from the standard thin disk model \citep{ShakuraSunyaev1973} for the same mass accretion rate in the following ways.
\begin{itemize}
    \item The disk is supported vertically by magnetic pressure rather than thermal pressure for these accretion rates. 
    
    \item The surface density and total optical depth are reduced by a factor of $\approx 10$. The disk scale height increases with radius significantly instead of being constant. 

	\item  A significant fraction of dissipation as well as the associated mass accretion (more than $50\%$ for $\dot{M}=0.07\Medd$) happen away from the disk midplane in the optically thin region, resulting in a high temperature corona inside $10r_g$. The fraction of dissipation in the corona region decreases with increasing mass accretion rate. 
	
	\item The transport of angular momentum is due to Maxwell and Reynolds stress inside the photosphere. But near the photosphere, radiation viscosity is the dominant mechanism for angular momentum transport, and this can dominate the vertically averaged stress.

\end{itemize}

\section*{Acknowledgements}
%We thank Geoffroy Lesur, Jeremy Goodman, Charles 
%Gammie, Julian Krolik as well as many participants of the accretion disk 
%program in KITP  for  helpful discussions. 
We thank Julian Krolik for helpful comments on an earlier draft. 
This research was supported in part by the National Science Foundation under Grant No. NSF PHY-1125915, 17-48958
and AST-1715277.
S.W.D. is supported by a Sloan Foundation Research Fellowship.
An award of computer time was provided by the Innovative and Novel Computational Impact on Theory and Experiment (INCITE) program. 
This research used resources of the Argonne Leadership Computing Facility, 
which is a DOE Office of Science User Facility supported under Contract DE-AC02-06CH11357.
Resources supporting this work were also provided by the NASA High-End Computing (HEC) 
Program through the NASA Advanced Supercomputing (NAS) Division at Ames Research Center. 

%\clearpage

%\end{thebibliography}

\bibliographystyle{astroads}
\bibliography{SubEddAGN}

\end{CJK*}

\end{document}